%% file: ms.tex
\documentclass{emulateapj}

\def\hd{\mbox{HD~160617}}
\def\hda{\mbox{HD~108317}}
\def\hdb{\mbox{HD~122563}}

\def\kmsec{\mbox{km~s$^{\rm -1}$}}
\def\logg{\mbox{log~{\it g}}}

\def\teff{\mbox{$T_{\rm eff}$}}
\def\vt{\mbox{$v_{\rm t}$}}
\def\rpro{\mbox{$r$-process}}
\def\spro{\mbox{$s$-process}}
\def\ncap{\mbox{$n$-capture}}

\slugcomment{Published in the Astrophysical Journal}
\shorttitle{Detection of All Three $r$-process Peaks}
\shortauthors{Roederer \& Lawler}

\begin{document}

\title{
Detection of Elements at All Three $r$-process Peaks \\
in the Metal-Poor Star HD 
160617\footnotemark[1]$^{~ \rm ,}$\footnotemark[2]$^{~ \rm ,}$\footnotemark[3]
}

\footnotetext[1]{
Some of the data presented in this paper were obtained from the 
Multimission Archive at the Space Telescope Science Institute (MAST). 
STScI is operated by the Association of Universities for Research in 
Astronomy, Inc., under NASA contract NAS5-26555. 
These data are associated with Program GO-8197.
}

\footnotetext[2]{
Based on data obtained from the European Southern Observatory (ESO) 
Science Archive Facility. 
These data are associated with Programs 65.L-0507 and 82.B-0610. 
}

\footnotetext[3]{
This research has made use of the Keck Observatory Archive (KOA), 
which is operated by the W.M.\ Keck Observatory and the NASA Exoplanet 
Science Institute (NExScI), under contract with the 
National Aeronautics and Space Administration.
These data are associated with Program H6aH (P.I.\ Boesgaard).
}

\author{
Ian U.\ Roederer\altaffilmark{4} and
James E.\ Lawler\altaffilmark{5}
}

\altaffiltext{4}{Carnegie Observatories, 
Pasadena, CA 91101, USA;
iur@obs.carnegiescience.edu
}
\altaffiltext{5}{Department of Physics, University of Wisconsin, 
Madison, WI 53706, USA;
jelawler@wisc.edu
}


\addtocounter{footnote}{5}

\begin{abstract}

We report the first detection of elements at all three
$r$-process peaks in the metal-poor halo star \mbox{HD~160617}.
These elements include arsenic and selenium,
which have not been detected previously in halo stars,
and the elements tellurium, osmium, iridium, and platinum, 
which have been detected previously.
Absorption lines of these elements are found in archive
observations made with the
Space Telescope Imaging Spectrograph onboard the
\textit{Hubble Space Telescope}.
We present up-to-date absolute atomic transition probabilities and 
complete line component patterns for these elements. 
Additional archival spectra of this star from several
ground-based instruments allow us to derive abundances or upper limits
of 45~elements in \mbox{HD~160617}, 
including 27~elements produced by neutron-capture reactions.
The average abundances of the elements at the three $r$-process peaks
are similar to the predicted solar system
$r$-process residuals when scaled to the 
abundances in the rare earth element domain.
This result for arsenic and selenium may be surprising
in light of predictions that the production of the
lightest $r$-process elements
generally should be decoupled from the
heavier $r$-process elements.

\end{abstract}


\section{Introduction}
\label{intro}

Understanding the origin of the elements is 
one of the major challenges of modern astrophysics. 
Heavy elements, here considered to be those more massive than the
iron-group, are primarily produced by neutron-capture (\ncap) reactions.
The heavy elements in the solar system (S.S.) were produced mainly by
two kinds of \ncap\ reactions, those slow (the \spro) or rapid (the \rpro)
relative to the average $\beta$-decay timescales of nuclei
along the reaction chain.
The nuclear structure of atoms and the conditions
at the time of nucleosynthesis determine the relative abundance patterns.
The abundance patterns can be interpreted through the lens of
experimental or theoretical properties of atomic nuclei to infer
the conditions present in the supernovae, merging neutron stars, or
red giant stars where these elements were forged.

Large energy gaps to the next highest nuclear energy level
occur in nuclei with
50, 82, and 126 neutrons.
Nuclei with these ``closed'' nuclear shells are especially stable and
resist further \ncap, so they are produced in greater abundance
during \ncap\ reactions.
The \spro\ path closely follows the valley of $\beta$-stability, whereas 
the \rpro\ blazes a path through neutron-rich nuclei.
The \rpro\ encounters the closed shells in nuclei of
lower proton number, $Z$, than the \spro\ does,
so the abundance peaks of the
\rpro\ are shifted to lower-mass atoms relative to the \spro\ peaks.

Most elements at the three \spro\ peaks are readily identified in
optical spectra obtained from ground-based instruments
(e.g., \citealt{merrill26,stjohn28,aller60,wallerstein63,vaneck01}).
These elements include
rubidium, strontium, yttrium, and zirconium (37~$\leq Z \leq$~40) 
at the first \spro\ peak;
barium, lanthanum, cerium, praseodymium, and neodymium (56~$\leq Z \leq$~60)
at the second \spro\ peak; and
lead and bismuth (82~$\leq Z \leq$~83)
at the third \spro\ peak.
In contrast, elements at the third \rpro\ peak---osmium,
iridium, and platinum (76~$\leq Z \leq$~78)---remained undetected
in halo stars until the Goddard High Resolution Spectrograph (GHRS) onboard
the \textit{Hubble Space Telescope} (\textit{HST}) revealed 
the ultraviolet (UV) spectra of late-type stars in high spectral resolution
\citep{cowan96,sneden98}.\footnote{
Numerous heavy elements, including those at the third \rpro\ peak,
have long been recognized in the stratified atmospheres of
chemically peculiar stars; e.g., \citet{morgan33} or \citet{dworetsky69}.
We do not regard these stars as useful for interpreting the
nucleosynthetic record, however, so we shall not discuss them further.
}
Subsequent investigations using the successor to the GHRS, the
Space Telescope Imaging Spectrograph (STIS),
continue to offer the only reliable window to detect these
third \rpro\ peak elements
\citep{cowan02,cowan05,sneden03,roederer09,roederer10a,roederer12b,barbuy11}.
Only one element at the second \rpro\ peak, tellurium ($Z =$~52),
was recently detected in three halo stars by \citet{roederer12a}.

The need for UV access can be explained using a combination of principles 
from astrophysics, nuclear physics, statistical mechanics, and atomic physics. 
Here we focus on photospheric temperatures typical for F, G, and K stars.
Many of the useful visible and UV lines of 
iron-group atoms and ions have excitation potentials (E.P.) 
of $\sim$~1~eV or more. 
Even the most abundant lighter \ncap\ elements tend to have 
abundances several orders of magnitude 
below those of iron-group elements. 
This simple fact eliminates the utility 
of atomic lines connecting highly excited levels of \ncap\ elements. 
It necessitates the use of resonance lines
(E.P.~$=$~0) or lines connected 
to low metastable levels if such levels exist 
in the atom or ion of interest. 
A simple consideration of Boltzmann factors 
reveals the ground and low metastable levels 
tend to serve as the major population reservoirs 
for both neutral and singly ionized atoms 
in a typical photosphere. 
Depending on the ionization potential (I.P.) of the neutral atom, 
either the neutral or singly ionized specie 
is dominant in F, G, and K stars.

Consider \ncap\ elements near and at 
the right edge of the periodic table. 
Rare gases at the right edge 
(krypton, $Z =$~36, and xenon, $Z =$~54) 
have deeply bound 
valence electrons due to their closed p-shells 
and a huge gap between the ground and lowest excited levels. 
The atomic structure of rare gases prohibits 
any photospheric abundance observations 
even on the strongly dominant neutral ionization stage,
whose resonance lines lie below 1500\,\AA\ 
(e.g., \citealt{morton00}).
The elements selenium ($Z =$~34) and bromine ($Z =$~35), 
adjacent to the rare gas krypton,
as well as tellurium and iodine ($Z =$~53),
adjacent to the rare gas xenon,
have partially filled p-shells. 
These have some similarities to the rare gases. 
There are gaps of $\sim$~48,000~cm$^{-1}$ and 67,000~cm$^{-1}$, 
respectively, between levels in the ground 4p$^{\rm n}$ configurations 
and the lowest opposite parity 4p$^{\rm n-1}$ 5s configurations 
of atomic selenium and bromine.
Even larger gaps exist in the singly charged ion of each element. 
The gaps between the ground 5p$^{\rm n}$ configurations 
and lowest opposite parity 5p$^{\rm n-1}$ 6s configurations 
of tellurium and iodine are somewhat smaller, 
but still greater than 44,000~cm$^{-1}$. 
Only UV lines with low E.P.\ values 
are even potentially useful for these elements.

Transition elements, such as osmium, iridium, and platinum,
with d-shell valence electrons tend to have lower I.P.\ values 
than elements with open p-shells. 
For many transition elements,
the singly-charged ion tends to be the dominant ionization stage. 
Spectral lines with low E.P.\ values almost always tend to occur 
at shorter wavelengths when one (or more) 
electron(s) is removed from an atom. 
As a result UV observations are needed 
for abundance studies on many transition elements.

The rare earth elements lanthanum through lutetium, 57~$\leq Z \leq$~71,
with f-shell valence electrons are important exceptions. 
Although the rare earth elements exist 
almost entirely as singly-ionized species, 
there are a great many visible transitions with low E.P.\ 
values due to the open f-shells
and interleaving low-lying 4f, 5p, and 6s orbitals. 
Their accessibility to ground-based astronomy 
provides a great opportunity for detailed studies 
of rare earth abundances (e.g., \citealt{sneden09}). 
The fact that rare earths are accessible to ground based astronomy 
provides some motivation for extensive lab work on these elements 
(e.g., \citealt{lawler09}). 
Alkaline earth elements, such as strontium ($Z =$~38) and barium ($Z =$~56),
with two valence s-shell electrons 
are also accessible to ground-based observations, 
even though they exist primarily as ionized species in typical photospheres.

Elements at the 
\rpro\ peaks are 
critical to understanding \rpro\ nucleosynthesis.
These serve as key normalization points 
to constrain the conditions that produce a
successful \rpro\ \citep{kratz93}.
Experimental data are available for 
only a limited number of short-lived nuclei along the 
\rpro\ path \citep{dillmann03,
baruah08}.
The Facility for Rare Isotope Beams, 
under construction at Michigan State University, 
and similar facilities in other countries 
will alleviate this shortage of \rpro\ nuclear data 
in the future.
Nuclei at the so-called ``waiting-points'' 
are the parent isobars of elements at the \rpro\ peaks,
including selenium and tellurium.
Yet, until now, the only secure measurements of their abundances
in environments suitable for inferring the \rpro\ abundance distribution
have come from the S.S.\ \rpro\ residuals.
This method of computing these residuals assumes that 
the \rpro\ produced some fraction of all heavy isotopes
in the S.S.\ except those attributed to \spro\ nucleosynthesis
or inaccessible to the \rpro\
(e.g., \citealt{seeger65,cameron82,kappeler89}).
Light \ncap\ nuclei may have a variety of origins besides
the $s$- and $r$-processes
(e.g., $\alpha$-rich freeze-out from nuclear statistical equilibrium; 
\citealt{woosley92}).
Since the assumptions underlying the S.S.\ \rpro\ residual distribution
for the light nuclei are tenuous,
detection of these elements in \rpro\
material beyond the S.S.\ is greatly desired.

Here, we report the detection of two light \ncap\ elements,
arsenic ($Z =$~33) and selenium in an archive STIS UV spectrum of the
bright, metal-poor subgiant \hd.
We also detect elements at the second and third \rpro\ peaks in \hd.
We supplement these data with ground-based archive spectra
to form a more complete estimate of the 
distribution of the heavy elements in \hd.
Since many of these transitions are rarely used for abundance 
analyses, we review the atomic physics literature and make
recommendations for the best available $\log(gf)$ values.
We also present line component patterns that account for
the hyperfine splitting (hfs) structure
present in some of these lines.

Throughout this work we use
the standard definitions of elemental abundances and ratios.
For element X, the logarithmic abundance is defined
as the number of atoms of element X per 10$^{12}$ hydrogen atoms,
$\log\epsilon$(X)~$\equiv \log_{10}(N_{\rm X}/N_{\rm H}) +12.0$.
For elements X and Y, the logarithmic abundance ratio relative to the
solar ratio of X and Y is defined as
[X/Y]~$\equiv \log_{10} (N_{\rm X}/N_{\rm Y}) -
\log_{10} (N_{\rm X}/N_{\rm Y})_{\odot}$.
Abundances or ratios denoted with the ionization state
indicate the total elemental abundance as derived from
that particular ionization state
after ionization corrections have been applied,
not the number density in that particular state.
When reporting relative abundance ratios for a specific element X
(e.g., [X/Fe]),
these ratios are constructed
by comparing the total abundance of element X derived from the neutral species 
with the total iron abundance derived
from Fe~\textsc{i} and the total abundance of element X derived from the
ionized species with the total iron abundance
derived from Fe~\textsc{ii}.

\section{Observations from the Archives}
\label{observations}

We use four sets of observations of \hd\ found in different archives.
The UV spectrum of \hd\
is taken from the 
Multimission Archive at the Space Telescope Science Institute.
These observations were made using STIS \citep{kimble98,woodgate98}
onboard \textit{HST}.
This spectrum has very high spectral resolution 
($R \equiv \lambda/\Delta\lambda =$~114,000)
and modest signal-to-noise (S/N) levels ranging from 
$\sim$~25--50 per pixel, but it
only covers a limited spectral range in the UV
(1879~$< \lambda <$~2148\,\AA).
We use the reduction and coaddition provided by the
StarCAT database \citep{ayres10}.

We use ground-based optical spectra of \hd\ from several
telescope and instrument combinations.
Some spectra were retrieved from the
European Southern Observatory (ESO) Science Archive Facility.
A spectrum obtained with the 
Ultraviolet and Visual Echelle Spectrograph (UVES; \citealt{dekker00}) 
on the Very Large Telescope (VLT) Kueyen at Cerro Paranal, Chile,
covers the wavelength range 3055--3870\,\AA\ 
at $R \sim$~40,000 
with S/N ranging from 100--400 per pixel.
A spectrum obtained with the 
High Accuracy Radial velocity Planet Searcher (HARPS; \citealt{mayor03}) 
on the ESO 3.6~m Telescope at La Silla, Chile,
covers the wavelength range 3780--6910\,\AA\ 
at $R \sim$~115,000
with S/N ranging from 150--300 per pixel.
Another spectrum was retrieved from the Keck Observatory Archive.
This spectrum was obtained with the
High Resolution Echelle Spectrometer (HIRES; \citealt{vogt94})
on the Keck~I Telescope at Mauna Kea, Hawai'i.
It covers the wavelength range 4395--6770\,\AA\
at $R \sim$~49,000
with S/N ranging from 130--300 per pixel.
These spectra were reduced and extracted
using the standard instrument pipelines before being downloaded, and
final processing was performed within the IRAF environment.

\subsection{Comparing Spectra Using Equivalent Width Measurements}

The different optical spectra of \hd\ overlap considerably in 
wavelength coverage.  
To assess their reliability, we compare the equivalent widths (EW)
of lines in common.
We measure the EWs using a semi-automatic routine that fits Voigt
absorption line profiles to continuum-normalized spectra.
We compare the EWs of species of lighter elements whose composition is
not expected to be
dominated by isotopes with hfs,
which could affect the line profiles.
These include
Mg~\textsc{i}, Si~\textsc{i}, 
Ca~\textsc{i}, Ti~\textsc{i} and \textsc{ii}, Cr~\textsc{i} and \textsc{ii},
Fe~\textsc{i} and \textsc{ii}, Ni~\textsc{i}, and Zn~\textsc{i}.
We find good internal agreement between the HARPS and HIRES spectra
($\langle\Delta\rangle = +$0.9~$\pm$~0.12~m\AA, $\sigma =$~1.6~m\AA, 174~lines).
The HARPS and UVES spectra have relatively little overlap, but even
there the agreement is reassuring
($\langle\Delta\rangle = -$1.0~$\pm$~1.6~m\AA, $\sigma =$~1.6~m\AA, 5~lines).
This suggests that
systematic differences are not apparent at a level greater than 1~m\AA,
so mixing abundances derived from different spectra should
not introduce any significant bias into the results.

The HARPS and HIRES spectra have comparable S/N levels, though the 
HARPS spectrum is of considerably higher spectral resolution.
The HIRES spectrum contains a few small order gaps redward of 5050\,\AA,
and the HARPS spectrum contains one 35\,\AA\ gap between the detectors
at 5305\,\AA.
Except when a line falls in this gap or appears distorted,
we prefer the abundances derived from the HARPS spectrum because
of the superior spectral resolution.

\section{Atomic Data}
\label{atomic}

We survey the literature on UV lines rarely used in abundance analyses, 
compare results from various experimental and theoretical studies, 
and identify the best available transition probabilities, hfs, and 
isotope shift (IS) data.
Here, we discuss these data in greater detail.

\subsection{Copper}

\input{tab1}

Seven lines of Cu~\textsc{ii}, all connecting metastable lower levels 
of the $^{1}$D and $^{3}$D terms in the 3d$^{9}$4s configuration 
to upper levels of the 3d$^{9}$4p configuration 
are of interest in this study. 
Due to the size of the spin-orbit splitting compared to term separations, 
both the upper and lower levels are best described by intermediate 
coupling schemes rather than by the LS scheme. 
In order of increasing energy, the LS terms 
of the upper configuration include $^{3}$P$^{\rm o}$, $^{3}$F$^{\rm o}$, 
$^{1}$F$^{\rm o}$, $^{1}$D$^{\rm o}$, $^{3}$D$^{\rm o}$, and $^{1}$P$^{\rm o}$, 
but levels with a common J in these terms are mixed. 
Level assignments for some levels in the compilation by 
\citet{sugar90} are not consistent with level assignments 
given by \citet{kono82}. 
Leading percentages given by \citet{donnelly99} 
are illustrative of the difficulty, 
for example the 3d$^{9}$4p $^{1}$D$^{\rm o}_{2}$ level 
(as assigned by \citeauthor{sugar90}) was found to be 
43\% 3d$^{9}$4p $^{3}$F$^{\rm o}_{2}$, 
24\% 3d$^{9}$4p $^{1}$D$^{\rm o}_{2}$, and 
19\% 3d$^{9}$4p $^{3}$D$^{\rm o}_{2}$. 
With intermediate coupling, only the parity and J selection rules 
for dipole transitions are consistently very strong. 
There has been extensive theoretical and experimental work on the 
3d$^{9}$4s--3d$^{9}$4p transition probabilities, 
but unfortunately there is significant scatter 
in the published results. 
Our preferred transition probabilities in Table~\ref{cutab} 
are based on experimental branching fractions from 
\citeauthor{kono82} and laser induced fluorescence (LIF) 
radiative lifetime measurements by \citet{pinnington97}
wherever available. 
Brief line-by-line discussions are appropriate in the case of Cu~\textsc{ii}.

The line at 2135\,\AA\ connects the upper 
3d$^{9}$4p $^{3}$F$^{\rm o}_{4}$ level to the lower 
3d$^{9}$4s $^{3}$D$_{3}$ level with a branching fraction of 1.0 
due to the J selection rule. 
\citet{pinnington97} reported a LIF measurement of 
2.18~$\pm$~0.04~ns for the upper level lifetime. 
Even if their $\sim$~2\% lifetime uncertainty is somewhat optimistic, 
the transition probability of the 2135\,\AA\ line 
is very well known. 
The line at 2126\,\AA\ connects the upper 
3d$^{9}$4p $^{3}$F$^{\rm o}_{2}$ level to the lower 
3d$^{9}$4s $^{3}$D$_{2}$ level 
with a branching fraction of 0.403 \citep{kono82}. 
This branching fraction has been confirmed to within 2\% by
\citet{crespo94}. 
Unfortunately no LIF lifetime measurement is available 
for the upper level. 
Inspection of Table~1 in \citeauthor{pinnington97}\
reveals that theoretical lifetimes by \citet{theodosiou86}
are generally in good agreement with the LIF measurements, 
and we combine the above experimental branching fraction with 
\citeauthor{theodosiou86}'s 2.34~ns 
theoretical lifetime to determine the transition probability 
for the 2126\,\AA\ line in Table~\ref{cutab}. 
No uncertainty is given in Table~\ref{cutab}, 
but it is likely smaller than $\pm$~15\%.

The line at 2112\,\AA\ connects the upper 
3d$^{9}$4p $^{1}$P$^{\rm o}_{1}$ to the lower 
3d$^{9}$4s $^{1}$D$_{2}$ level, 
and this is the most problematic case of Table~\ref{cutab}. 
The difficulty is due to the strong 1358\,\AA\ line 
from the upper $^{1}$P$^{\rm o}_{1}$ level 
to the ground 3d$^{10}$ $^{1}$S$_{0}$ level. 
There is no direct experimental measurement of the 
1358\,\AA\ branching fraction, but the short upper level 
lifetime of 1.34~$\pm$~0.22~ns from \citet{pinnington97} 
and elementary theoretical considerations 
indicate that the line has a significant branching fraction. 
\citet{brown09} used the beam foil technique to confirm 
the short upper level lifetime 
and have recommended a branching fraction of 
0.45 for the 1358\,\AA\ line based on earlier theoretical work. 
By combining this branching fraction for the 1358 Å line, 
measurements of the branching ratios of other emission lines 
from the upper $^{1}$P$^{\rm o}_{1}$ level by 
\citet{kono82}, and the LIF lifetime measurement by \citeauthor{pinnington97}, 
we compute a transition probability for the 
2112\,\AA\ line given in Table~\ref{cutab}.
No uncertainty can be given for the 
2112\,\AA\ transition probability without additional experimental work.

The next two lines in Table~\ref{cutab} at 2104\,\AA\ and 2054\,\AA\ 
are both from the upper 3d$^{9}$4p $^{1}$D$^{\rm o}_{2}$ level 
(assigned as $^{3}$D$^{\rm o}_{2}$ by \citealt{kono82}) 
to the lower 3d$^{9}$4s $^{3}$D$_{1}$ and $^{3}$D$_{2}$ levels 
with measured branching fractions of 0.229 and 0.402 respectively. 
A LIF lifetime measurement of 2.48~$\pm$~0.05~ns by \citet{pinnington97}
is available for the upper level. 
Uncertainties on the transition probabilities are from the 
2\% lifetime uncertainty and from a 7\% branching fraction uncertainty for
each line. 
The 2037\,\AA\ from the upper 3d$^{9}$4p $^{3}$D$^{\rm o}_{3}$ level 
(assigned as $^{1}$F$^{\rm o}_{3}$ by \citeauthor{kono82}) 
to the lower 3d$^{9}$4s $^{3}$D$_{2}$ level 
also has a measured branching fraction of 0.296 
from \citeauthor{kono82} and a LIF lifetime measurement of 
2.21~$\pm$~0.04~ns by \citeauthor{pinnington97}
This branching fraction of the 2037\,\AA\ line 
was confirmed to within 2\% by \citet{crespo94}. 
The last line in Table~\ref{cutab} at 1979\,\AA\ 
connects the upper 3d$^{9}$4p $^{3}$D$^{\rm o}_{2}$ level 
(assigned as $^{1}$D$^{\rm o}_{2}$ by \citeauthor{kono82}) 
to the lower 3d$^{9}$4s $^{3}$D$_{2}$ level 
with a measured branching fraction of 0.222. 
The upper level has a LIF lifetime measurement of 
2.51~$\pm$~0.09~ns by \citeauthor{pinnington97}\
is combined with the branching fraction to determine 
the transition probability in Table~\ref{cutab}. 

Energy levels in Table~\ref{cutab}
are from the compilation by \citet{sugar90}
based on interferometric measurements by 
\citet{reader60}. 
Air wavelengths for lines above 2000\,\AA\ 
are from the level energies and index of air \citep{peck72}.

The Cu~\textsc{ii} lines of interest are expected to have 
non-negligible hfs and IS structure because 
Cu is an odd $Z$ element with two I~$=$~3/2 isotopes. 
The S.S.\ isotopic abundances are 69.15\% for $^{63}$Cu and 
30.85\% for $^{65}$Cu \citep{bohlke05}. 
These isotopes have nearly equal nuclear magnetic dipole moments 
from the unpaired proton in each nucleus. 
The only available experimental hfs and IS data 
for the lines of interest are those reported by \citet{elbel63}, 
and unfortunately these are not accurate by modern 
(laser spectroscopy) standards. 
\citet{elbel69} later reported some theoretical work on the 
IS of the 3d$^{9}$4s levels. 
We use the available experimental data on the hfs and IS data from 
\citeauthor{elbel63}\ for the 3d$^{9}$4s and 3d$^{9}$4p 
levels to generate complete line component patterns of the 
strongest Cu~\textsc{ii} lines of interest.
We find that the abundance results for the 
2126\,\AA\ and 2112\,\AA\ lines are changed by less than 0.01~dex
when including the effect of hfs.
Complete line component patterns for Cu~\textsc{ii} lines 
are omitted from this paper 
with the expectation that more accurate lab data 
will soon be available for the Cu~\textsc{ii} lines of interest.

\subsection{Zinc}

\input{tab2}

Lines of both Zn~\textsc{i} and Zn~\textsc{ii} are of interest in this study. 
The multiplet of Zn~\textsc{i} near 4750\,\AA\ 
connecting the 
lower 4s4p $^{3}$P$^{\rm o}$ term 
to the 
upper 4s5s $^{3}$S$_{1}$ level 
is a nearly pure Russell-Saunders (LS) multiplet. 
Early experimental branching fractions 
\citep{schuttevaer43} as well as theoretical considerations 
are consistent with the nearly pure LS character. 
For the lower 4s4p $^{3}$P$^{\rm o}$ term the relatively small 
fine structure splitting, $<$~600~cm$^{-1}$, 
compared to the separation, $>$~14000~cm$^{-1}$, 
from other levels which can mix is indicative of a nearly pure LS term. 
Several independent measurements and calculations 
of the upper level lifetime, including the preferred 
LIF measurement of 8.0(4)~ns 
\citep{kerkhoff80}, are consistent to $\pm$~10\%. 
The selected transition probabilities in Table~\ref{zntab} 
are from \citeauthor{kerkhoff80}\
The second multiplet of Zn~\textsc{i} near 3300\,\AA\ 
connecting the 
same lower 4s4p $^{3}$P$^{\rm o}$ term 
to the 
upper 4s4d $^{3}$D term 
is also a nearly pure LS multiplet. 
Similar to the above case the early experimental branching fractions 
\citep{schuttevaer43} as well as theoretical considerations 
are consistent with the nearly pure LS character. 
For the upper 4s4d $^{3}$D term 
the very small fine structure splitting, 
$<$~10~cm$^{-1}$ total, 
compared to the separation, $>$~300~cm$^{-1}$, 
from other levels which can mix 
is indicative of a nearly pure LS term. 
Several independent sets of measurements and calculations 
of the upper level lifetimes, 
including two sets of LIF measurements 
\citep{kerkhoff80,blagoev04}, are consistent 
to within $\pm$~10\% for all three levels of the upper term. 
The selected results in Table~\ref{zntab} are from 
\citeauthor{kerkhoff80}
These values agree very well the critical compilation by 
\citet{reader80}.

The very weak spin-forbidden line at 3075\,\AA\ 
connects the ground 4s$^{2}$ $^{1}$S$_{0}$ 
levels of the neutral atom to the 4s4p $^{3}$P$^{\rm o}_{1}$ level. 
This weak line is useful in this abundance study 
due to the relatively high abundance of zinc 
and due to the ionization balance 
only slightly favoring the singly ionized specie over the neutral atom. 
The branching fraction of the 3075\,\AA\ line is 1.0;
however, the radiative lifetime is too long 
for standard LIF experimental setups. 
This is expected because the 4s4p $^{3}$P$^{\rm o}$ term is nearly pure LS 
and the 4s4p $^{3}$P$^{\rm o}_{1}$ level 
is only very slightly mixed with the much higher 
4s4p $^{1}$P$^{\rm o}_{1}$ level 
yielding a radiative lifetime longer than 10~$\mu$s. 
The critical compilation of \citet{reader80} 
included a transition probability with an accuracy B 
($\pm$~10\% or $\pm$~0.04~dex),
but recent theoretical values have diverged somewhat 
from the 1980 compiled result of $\log(gf) = -$3.854. 
\citet{chen10} found $\log(gf) = -$3.72. 
\citet{liu06} found $-$3.78. 
\citet{chou94} found $-$3.89. 
For the near future we recommend that the compilation value of 
$\log(gf) = -$3.854 be used, 
but with the realization the uncertainty is somewhat larger than 0.04~dex.

Only the 2062.00\,\AA\ line of Zn~\textsc{ii} is included in this study. 
This line has a branching fraction of 1.0 and connects the ground 
4s $^{2}$S$_{1/2}$ level to the 4p $^{2}$P$^{\rm o}_{1/2}$ level. 
\citet{bergeson93} measured the upper level lifetime using LIF. 
More recent theoretical investigations have reproduced their 2.5(2)~ns 
radiative lifetime measurement to within one-half of an error bar 
\citep{harrison03,dixit08a}.
The $\log(gf)$ in Table~\ref{zntab} is the value from their LIF experiment.

New lab measurements using Fourier transform spectrometers of both 
Zn~\textsc{i} and Zn~\textsc{ii} 
wavenumbers or wavelengths have recently been performed 
to support studies of possible changes in the fine structure constant 
during the expansion of the Universe 
\citep{gullberg00,pickering00}.
Energy levels in Table~\ref{zntab}
are from \citeauthor{gullberg00}.
The Zn~\textsc{ii} wavelengths from 
\citeauthor{pickering00}\ are in perfect (0.0001\,\AA) 
agreement with those of \citeauthor{gullberg00}.
Wavelengths are computed from the energy levels 
using the index of air \citep{peck72}.

\subsection{Arsenic}

\input{tab3}

\input{tab4-stub}

Three of the four As~\textsc{i} lines of interest in this study 
are members of a multiplet between the ground 
4s$^{2}$4p$^{3}$ $^{4}$S$^{\rm o}$ term and the excited 
4s$^{2}$4p$^{2}$($^{3}$P)5s $^{4}$P term. 
The fourth line at 1890\,\AA\ is from the metastable 
4s$^{2}$4p$^{3}$ $^{2}$D$^{\rm o}_{3/2}$ level 
of the ground configuration to the excited 
4s$^{2}$4p$^{2}$($^{1}$D)5s $^{2}$D$_{3/2}$ level. 
There has been both experimental and theoretical work 
on the transition probabilities of these 4p--5s lines. 
\citet{bengtsson92a} reported radiative lifetime measurements 
using LIF on two of the three levels 
in the 4s$^{2}$4p$^{2}$($^{3}$P)5s $^{4}$P term, 
and they used these to normalize experimental branching fractions 
measured by \citet{lotrian80}
for the 1972\,\AA\ and 1937\,\AA\ lines
included in Table~\ref{astab}. 
Russell-Saunders (LS) coupling is fairly good 
for these low levels of As, 
and the lines of interest are dominant branches,
which minimizes the branching fraction uncertainty. 
\citet{holmgren75} and later \citet{bieron92} 
computed transition probabilities for As~\textsc{i} lines of interest
and found good agreement with the above experimental results.
\citeauthor{holmgren75}'s theoretical transition probabilities 
for the 1990\,\AA\ and 1890\,\AA\ lines are 
included in Table~\ref{astab}
with experimental results for the other two lines.

Arsenic has one stable isotope with a nuclear spin of I~$=$~3/2. 
Hyperfine patterns for the lines of interest are provided in 
Table~\ref{ashfstab}, based on extremely accurate 
hyperfine A and B coefficients for the ground level.
These were 
measured using a radio frequency technique by 
\citet{pendlebury64}, and hyperfine A coefficients 
for other levels were measured 
using an optical technique by \citet{bouazza87}.
The vacuum wavelengths and energy levels are from \citet{howard85}.

\subsection{Selenium}

\input{tab5}

Three of the four Se~\textsc{i} lines of interest in this study 
are members of a multiplet between the ground 
4s$^{2}$4p$^{4}$ $^{3}$P term and the excited 
4s$^{2}$4p$^{3}$($^{4}$S)5s $^{3}$S$^{\rm o}_{1}$ level. 
The fourth line is spin forbidden and thus weak. 
It connects the ground $^{3}$P$_{2}$ level to the excited 
4s$^{2}$4p$^{3}$($^{4}$S)5s $^{5}$S$^{\rm o}_{2}$ level. 
After careful review of both older and more recent literature, 
we conclude that \citeauthor{morton00}'s (2000) recommendations 
for wavelengths and transition probabilities 
of these Se~\textsc{i} lines are still up-to-date. 
\citeauthor{morton00}\ chose radiative lifetimes 
measured using LIF by 
\citet{bengtsson92b,bengtsson92c}
for both upper levels and combined these with emission branching fractions 
from \citet{ubelis86}.
\citeauthor{morton00}\ used energy levels from 
\citet{morillon74} with a correction of $+$0.23~cm$^{-1}$ 
for higher levels recommended by \citet{lindgren77}.
These results are summarized in Table~\ref{setab}.

There are 6 naturally-occurring isotopes of selenium found on earth.
The isotope shifts are small and can be neglected for our purposes.
Only one isotope, $^{77}$Se, has non-zero nuclear spin I~$=$~1/2.
This isotope comprises only 7.6\% of natural selenium
\citep{bohlke05}, so its 
hfs can also be neglected for our purposes.

\subsection{Molybdenum}

\input{tab6}

Four lines of Mo~\textsc{ii}, 
all connected the ground 4d$^{5}$ a$^{6}$S$_{5/2}$ level, 
are of interest in this study. 
Three of the four upper levels of interest belong to the 
4d$^{4}$($^{5}$D)5p z$^{6}$P$^{\rm o}$ term. 
The fourth and highest upper level is now assigned as the 
4d$^{4}$($^{5}$D)5p z$^{4}$D$^{\rm o}_{5/2}$ level, 
but it is significantly mixed with the 
4d$^{4}$($^{5}$D)5p z$^{6}$P$^{\rm o}_{5/2}$ level \citep{nilsson03}. 
Publications prior to the 2003 reanalysis of Mo~\textsc{ii}
by \citeauthor{nilsson03}\ have different assignments 
for the highest and lowest levels of Table~\ref{motab}
and have less accurate level energies than achieved in 
2003 using a Fourier transform spectrometer. 
In addition to the transitions to the ground level, 
these four upper levels have strong branches to metastable levels 
in the 4d$^{4}$($^{5}$D)5s a$^{6}$D term. 
Experimental branching fractions reported by 
\citet{sikstrom01} are preferred 
and are in a satisfactory agreement with theoretical values 
reported in the same paper for three of the four lines of interest. 
Radiative lifetimes of the four upper levels 
are all in the few nanosecond range due to the UV wavelengths 
of the highly allowed emission lines. 
The earlier LIF measurements of the radiative lifetimes of interest 
by \citet{hannaford83} do not agree with the 
LIF measurements by \citeauthor{sikstrom01}
Differences range from 18\% to 38\%, 
which is unusual for two sets of LIF measurements. 
Theoretical lifetimes reported by \citeauthor{sikstrom01}\
and by \citet{lundberg10} are in agreement 
with the experimental lifetimes of \citeauthor{sikstrom01}
The primary cause of the discordant results between the two LIF experiments 
is very likely limitations of the 
fluorescence detection system used by \citeauthor{hannaford83}.
Accurate LIF measurements on lifetimes ranging down to $\sim$~2~ns 
are possible using analogue time-resolved fluorescence detection methods, 
but some precautions are necessary. 
It is essential to establish that the electronic bandwidth, 
linearity, and overall fidelity of the electronic detection system 
including the photomultiplier, cables, and digitizer 
(or similar recording system) are adequate for 2~ns lifetime measurements. 
This is best accomplished by measuring a very short benchmark lifetime 
such as that of the resonance level of Be~\textsc{ii}, 
which is very well known from ab-initio theory \citep{weiss95}. 
Tests of the combined electronic detection system 
are much better than tests on individual parts at GHz 
electronic frequencies. 
The selected transition probabilities of Table~\ref{motab}
are the experimental values of \citeauthor{sikstrom01}
Table~\ref{motab} has up-to-date level energies and air wavelengths
\citep{peck72,nilsson03}.

\subsection{Cadmium}

\input{tab7}

\input{tab8-stub}

Only the Cd~\textsc{ii} resonance doublet from the ground 
4d$^{10}$ 5s $^{2}$S$_{1/2}$ level to the 
4d$^{10}$ 5p $^{2}$P$^{\rm o}$ term 
is useful for stellar abundance studies. 
Other lines are at wavelengths that are too short or have 
E.P.'s that are too high, 
except possibly in very hot stars. 
The branching fractions of these UV resonance lines are 
1.0 and the radiative lifetimes are known to exquisite accuracy and precision 
($\sim$~0.4\%) 
from LIF measurements by \citet{moehring06} and from earlier, 
but somewhat less accurate, measurements by others.
(See references in \citealt{moehring06}.)

There are six even isotopes with no nuclear spin 
and two odd isotopes with I~$=$~1/2 and significant ($>$~12\%) 
relative abundances in natural cadmium
\citep{bohlke05}.
The unpaired s-electron in the ground level of this ion 
yields wide hyperfine structure (hfs) in the odd isotopes 
and significant isotopic structure from field shifts 
(nuclear volume effect). 
The hfs A value for the ground level of $^{113}$Cd$^{+}$ 
is known to exquisite accuracy and precision 
from radio frequency measurements by \citet{jelenkovic06}.
The selected hfs A values of the excited levels of $^{113}$Cd$^{+}$ 
are from calculations of \citet{dixit08b}.
There are older measurements of the excited level hfs A values by 
\citet{brimicombe76}
that agree with the theoretical values from \citeauthor{dixit08b}\
within 0.001~cm$^{-1}$. 
We use theoretical hfs A values of \citeauthor{dixit08b}\
in part because their calculated ground level hfs A value was in 
much better agreement with very accurate 
and precise radio frequency measurements than was 
the experimental result from \citeauthor{brimicombe76}\

The ratio of the hfs A values for $^{111}$Cd$^{+}$ to that of 
$^{113}$Cd$^{+}$ is from \citet{spence72}.
Isotope shifts are from \citet{brimicombe76}. 
Energy levels are from \citet{burns56},
and the index of air is from \citet{peck72}.
Tables~\ref{cdtab} and \ref{cdhfstab}
summarize the selected lab data for the Cd~\textsc{ii}
resonance lines. 
S.S.\ isotopic abundances are used in Table~\ref{cdhfstab}
\citep{bohlke05}.

\subsection{Ytterbium}

\input{tab9}

\input{tab10-stub}

The two UV Yb~\textsc{ii} lines of interest in this study 
connect the ground 4f$^{14}$6s $^{2}$S$_{1/2}$ level to a 
J~$=$~3/2 level at 47005.46~cm$^{-1}$ and a 
J~$=$~1/2 level at 47228.96~cm$^{-1}$, 
both with 4f$^{13}$($^{2}$F$^{\rm o}_{5/2}$)5d6s($^{1}$D) assignments. 
These high lying levels are not pure, 
but both UV lines are strongly dominant 
($>$~0.9) branches. 
There is a LIF measurement of the radiative lifetime of the 
47005~cm$^{-1}$ level \citep{pinnington94}. 
The measured branching fraction for the decay of the 
47005~cm$^{-1}$ level at 2126\,\AA\ to the ground level is 0.988 
\citep{kedzierski10}.
The transition probability for the 2126\,\AA\ line 
from the modern LIF lifetime measurement 
and recent branching fraction measurement 
agrees beautifully with an earlier quantum calculation 
\citep{biemont98}.
Without a modern experimental value 
for the transition probability of the 2116\,\AA\ line, 
it is best to use the theoretical transition probability from 
\citeauthor{biemont98}
Energy levels are from \citet{martin78}, and 
wavelengths are computed using the index of air \citep{peck72}.
Table~\ref{ybtab} summarizes these data. 

There are 5 stable even isotopes of ytterbium with no nuclear spin 
and 2 odd isotopes, 
$^{171}$Yb with I~$=$~1/2 and 
$^{173}$Yb with I~$=$~5/2, 
both with significant ($>$~14\%) 
relative abundance in natural ytterbium \citep{bohlke05}. 
The ground level has large hfs due to the unpaired s-electron, 
and fortunately hfs A values for the odd isotopes are well known 
\citep{munch87,casdorf91}. 
Unfortunately no hfs data are available 
on the excited levels at 47005~cm$^{-1}$ and 47228~cm$^{-1}$. 
\citet{ahmad97} reported experimental level isotope shifts for 
$^{172}$Yb$^{+}$ and $^{176}$Yb$^{+}$ 
that can be used to determine the 2126\,\AA\ and 2116\,\AA\ 
isotope shift of at least these two isotopes. 
We have extended their measurements to other isotopes 
using a modified King plot as suggested by 
\citet{theodossiou97}.
A rough estimate is used for the $^{168}$Yb isotope 
that has a minor (0.13\%) relative abundance. 
Table~\ref{ybhfstab} presents hfs and S.S.\ isotopic line component patterns 
for the lines of interest, 
but the reader is advised that the hfs A 
of upper levels of the odd isotopes have been neglected.

\subsection{Platinum}

\input{tab11}

Two of three UV Pt~\textsc{i} lines of interest in this study 
connect the ground 4d$^{9}$6s $^{3}$D$_{3}$ level to the 
5d$^{8}$6s6p($^{4}$F) $^{3}$F$^{\rm o}_{4}$ level at 48351.94~cm$^{-1}$ and the 
5d$^{8}$6s6p($^{4}$F) $^{3}$D$^{\rm o}_{3}$ level at 48779.337~cm$^{-1}$. 
The third line connects the low metastable 5d$^{8}$6s$^{2}$ $^{3}$F$_{4}$ 
level at 823.678~cm$^{-1}$ to the 
5d$^{8}$6s6p($^{4}$F) $^{3}$F$^{\rm o}_{4}$ level. 
The radiative lifetimes of both upper levels were measured 
using LIF \citep{denhartog05}. 
Branching fractions and absolute transition probabilities 
were also measured for the lines at 2103\,\AA\ and 2067\,\AA\ 
connected to the 5d$^{8}$6s6p($^{4}$F) $^{3}$F$_{4}$ level by 
\citeauthor{denhartog05}
A branching fraction measurement for the third line at 2049\,\AA\ 
was reported by \citet{lotrian82}.
We used this measurement and the
lifetime measured by \citeauthor{denhartog05}\ 
to determine its transition probability, as given in Table~\ref{pttab}.
Energy levels in Table~\ref{pttab} 
are from \citet{blaise92},
and air wavelengths are computed from the levels 
using the index of air \citep{peck72}.

\subsection{Mercury}

\input{tab12-stub}

Only the resonance line of Hg~\textsc{ii} at 1942\,\AA\ 
is potentially useful for abundance determinations 
in most F, G, and K stars. 
This line connects to the ground 5d$^{10}$6s $^{2}$S$_{1/2}$ level 
to the 5d$^{10}$6p $^{2}$P$^{\rm o}_{1/2}$ level 
with a branching fraction of 1.0. 
The best available radiative lifetime measurement of 
2.91~$\pm$~0.11~ns is by \citet{pinnington88}
using the Beam Foil technique with an 
Arbitrarily Normalized Decay Curve (ANDC) analysis. 
Although early Beam Foil measurements 
were often inaccurate due to the non-selective excitation 
of the method and resultant cascade repopulation, 
this problem is eliminated using an ANDC analysis. 
The experimental $\log(gf)$ from \citeauthor{pinnington88}\
is $-$0.410. 
A series of sophisticated quantum calculations 
have also been published during the last decade or so 
using a variety of methods. 
Although a single valence s-electron or p-electron
outside of a closed d-shell might appear 
to be straightforward from a theoretical perspective, 
one must keep in mind that core polarization 
can be important, and relativistic effects are significant 
in an atom or ion as heavy as Hg. 
\citet{brage99} used a fully relativistic multiconfiguration 
Dirac-Fock method and found $\log(gf) = -$0.418. 
\citet{safronova04} used relativistic many-body perturbation theory 
and found $\log(gf) = -$0.369. 
\citet{glowacki09} used a configuration-interaction method 
with numerical Dirac-Fock wave functions and found $\log(gf) = -$0.381. 
Most recently, \citet{simmons11} used a variety 
of multi-order Coulomb as well as Coulomb-Breit approximations 
and found $\log(gf) = -$0.363. 
Based on the above experimental and theoretical results 
we recommend $\log(gf) = -$0.40~$\pm$~0.04, 
along with a transition wavenumber of 
51486.070~cm$^{-1}$ and vacuum wavelength of 
1942.2729\,\AA\ for the S.S.\ 
isotopic mix from \citet{sansonetti01}. 
We note that our recommended $\log(gf)$ does not agree with 
the value from \citeauthor{sansonetti01} 
that is currently in the 
Atomic Spectra Database.\footnote{
\url{http://www.nist.gov/pml/data/asd.cfm}
}

The wide hfs and S.S.\ isotopic structure 
of the 1942\,\AA\ line 
was measured quite well by \citet{guern77}. 
An incredibly accurate and precise 
(uncertainty $<$~0.001~Hz) 
radio frequency measurement of the $^{199}$Hg$^{+}$ ground level hfs splittings 
was published by \citet{berkeland98}, 
motivated by atomic clock research.
\citet{burt09} reported a similar measurement for 
$^{201}$Hg$^{+}$ ground level hfs splitting. 
These improved hfs A values are used 
to update values from \citeauthor{guern77}\ in Table~\ref{hghfstab}, 
but the updates shift most hfs components by $<$~0.001~cm$^{-1}$. 
We estimate the isotope shift for $^{196}$Hg,
which has a low abundance of 0.15\% in S.S. material.
For purposes of stellar abundance analysis,
the line component pattern presented in Table~\ref{hghfstab}
is equivalent to that presented by
\citet{leckrone91}. 

\section{Abundance Analysis}
\label{analysis}

\subsection{Model Atmospheres}

\hd\ lies relatively near the Galactic midplane ($b = -$5$^{\circ}$).
Its parallax, measured by the Hipparcos mission (using the
reduction validated by \citealt{vanleeuwen07}),
places it at a distance of 110~$\pm$~20~pc, which is
within the reddening layer.
The reddening at infinity along the line of sight to \hd\
predicted by all-sky dust maps (e.g., \citealt{schlegel98}),
$E(B-V) =$~0.53, almost certainly overestimates
the reddening between \hd\ and the Sun.
This interpretation finds support in the
low reddening value derived from Str\"{o}mgren photometry,
$E(b-y) =$~0.011 \citep{nissen02},
as kindly pointed out by the referee.
It is inadvisable in this case to derive model atmosphere parameters
by photometric methods (e.g., color-temperature relations).
Instead, we derive these parameters from the spectrum of \hd\
and comparison with a much larger sample of 153 metal-poor field subgiants
from an unpublished study (I.U.\ Roederer et al., in preparation).
This sample does not include \hd.
We interpolate model atmospheres from the grid of \citet{castelli03}.
We perform the abundance calculations using the latest version
of MOOG \citep{sneden73}, which includes the contribution of
Rayleigh scattering from atomic H~\textsc{i}
in the source function \citep{sobeck11}.
For this warm atmosphere, we find
no difference ($<$~0.01~dex) in the abundances derived using
earlier versions of MOOG,
even for lines with wavelengths short of 2000\,\AA.

For the large, unpublished sample, we adopt an approach using
broadband colors (usually $V-K$) and 12~Gyr isochrones to make an
initial estimate of the effective temperature, \teff, and 
(logarithm of the) surface gravity, \logg.
These values are then refined iteratively until there is no 
trend of Fe~\textsc{i} abundance versus lower E.P.\
(used to derive \teff)
or line strength (used to derive the microturbulent velocity, \vt).
For stars near 6000~K, that study finds a mean difference in the 
iron abundance 
derived from Fe~\textsc{ii} and Fe~\textsc{i} of about $+$0.15~dex
with a standard deviation of about 0.10~dex.
This matches the approximate offset expected 
for stars at similar stellar parameters 
(\teff~$\approx$~6000~K) and metallicities
([Fe/H]~$\approx -$1.8) if departures from 
local thermodynamic equilibrium (LTE) are responsible
(e.g., \citealt{thevenin99,mashonkina11}).

In the present study, we set \teff\ by requiring no trend of 
derived Fe~\textsc{i} abundance with E.P., and we set \vt\ by
requiring no trend with line strength.
Based on the large sample of metal-poor subgiants, 
we set \logg\ by requiring 
$\log \epsilon$~(Fe~\textsc{ii})~$= \log \epsilon$~(Fe~\textsc{i})~$+$~0.15.
We set the overall metallicity of the model atmosphere equal to the
iron abundance derived from Fe~\textsc{ii}.
We iteratively remove lines from the input list if their calculated
abundance deviates from the mean by more than 2$\sigma$.

For \hd,
we derive \teff~$=$~5950~K, \logg~$=$~3.90, \vt~$=$~1.3~\kmsec, and
[M/H]~$= -$1.78.
For the subgiants in the large sample, we find typical
internal (i.e., statistical) uncertainties to be
$\sigma_{T} =$~46~K,
$\sigma_{\logg} =$~0.18,
$\sigma_{vt} =$~0.06~\kmsec, and
$\sigma_{\rm [M/H]} =$~0.09.
The absolute uncertainties are much more difficult to assess.
Through comparisons with previous studies of the same stars 
in that sample (40~stars), 
we estimate the absolute uncertainties may be as large as
$\sigma_{T} =$~200~K,
$\sigma_{\logg} =$~0.35,
$\sigma_{vt} =$~0.35~\kmsec, and
$\sigma_{\rm [M/H]} =$~0.2.
We adopt these representative uncertainties for \hd.
Our adopted model parameters are in good 
(1$\sigma$ internal) agreement with the means of recent 
studies of \hd:
$\langle$\teff$\rangle$~$=$~5965~K ($\sigma =$~49~K),
$\langle$\logg$\rangle$~$=$~3.75 ($\sigma =$~0.09),
$\langle$\vt$\rangle$~$=$~1.44~\kmsec\ ($\sigma =$~0.12~\kmsec),
$\langle$[Fe/H]$\rangle$~$= -$1.74 ($\sigma =$~0.12)
\citep{thevenin99,gratton00,nissen02,nissen07,akerman04,caffau05,jonsell05,
asplund06,johnson07,tan09,hansen11,peterson11}.

\subsection{Derivation of Abundances}

\input{tab13-stub}

\input{tab14-mystub}

\input{tab15}

We perform a standard EW abundance analysis 
for unblended lines with negligible hfs.
The EWs for these lines are listed in Table~\ref{ewtab}.
For all other lines we derive abundances by comparing the 
observed line profiles with synthetic spectra
computed using MOOG.
In all cases except copper we adopt the predicted \rpro\ isotopic distribution
as given in \citet{sneden08}.
For copper we adopt the S.S.\ isotopic distribution 
\citep{bohlke05}.

Abundances for all 377~lines examined are presented in Table~\ref{linetab}.
Abundances that are less secure are indicated by a ``:'' and 
given half weight in the final averaging.
We assume a minimal internal uncertainty of 0.10~dex per line.
Final mean abundances for \hd\ are presented in Table~\ref{abundtab}.

\subsection{Are We Detecting the Majority Species of These Elements?}
\label{saha}

\begin{figure*}
\begin{center}
\includegraphics[angle=0,width=4.8in]{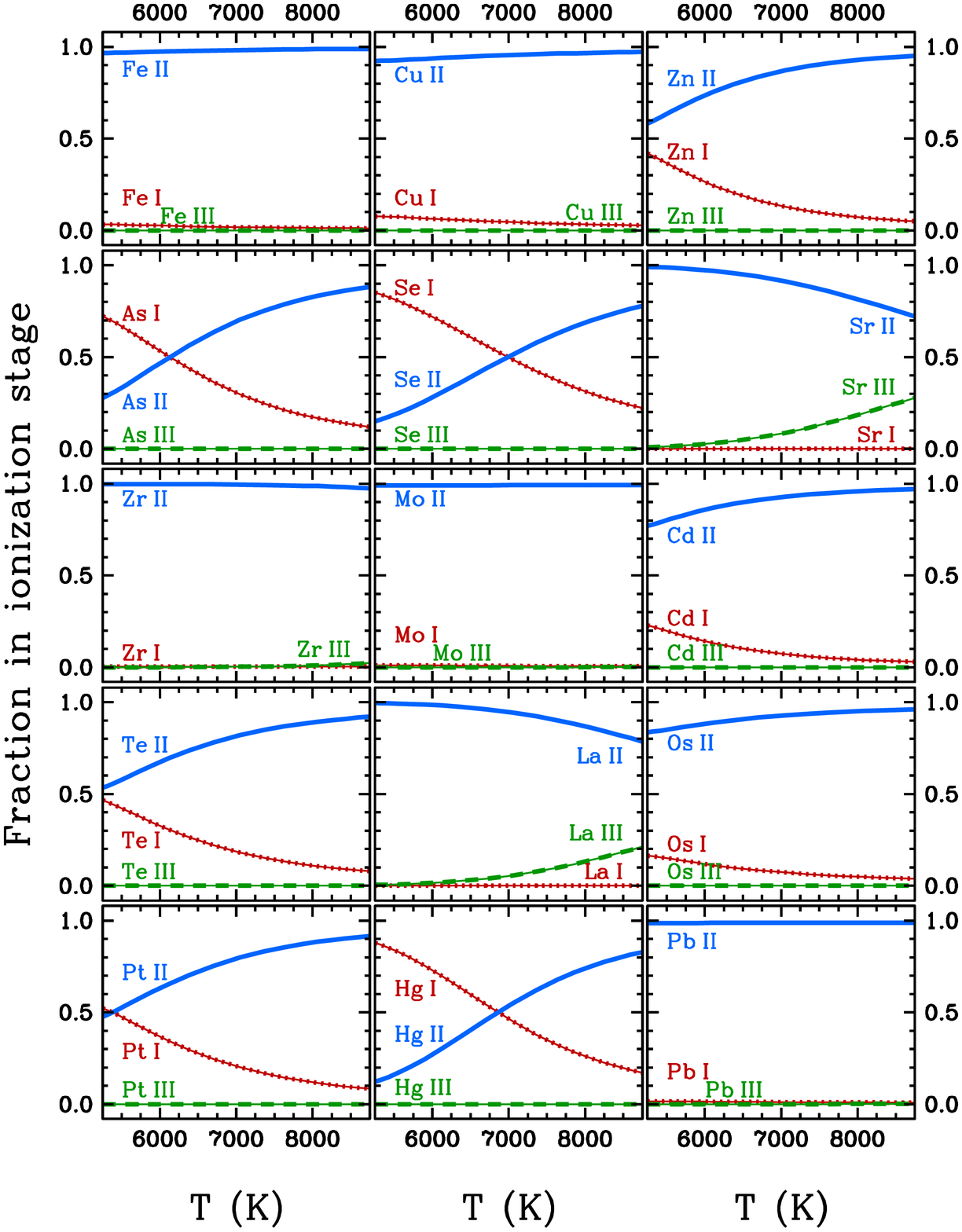}
\end{center}
\caption{
\label{sahaplot}
Illustration of the ionization distribution for 15~elements
of interest in the atmosphere of \hd.
The fraction of an element in the neutral state is shown by the
red studded curves.
The fraction in the first ionized state is shown by the
blue solid curves.
The fraction in the second ionized state is shown by the 
green dashed curves.
The temperature range shown corresponds to the main line-forming layers
for weak lines
($\log$~EW/$\lambda$~$< -$4.7) in the atmosphere of \hd.
For weak lines with low E.P.\ values,
$>$99\% of the line opacity 
is found between 5500~K and 8500~K 
in this model atmosphere.
 }
\end{figure*}

In an LTE abundance analysis, we assume that the distribution of
atoms among different level populations and ionization states
can be determined by knowing the temperature and pressure (or density)
at a particular layer in the atmosphere.
These abundances may be in error if Boltzmann or Saha
equilibrium is not satisfied.
Ionization corrections for minority species account for a 
proportionally larger fraction of the line opacity, so
abundances derived from minority species of a given element
may be particularly sensitive to departures from LTE.

Several of the elements considered in the present study are
rarely examined in late-type stellar atmospheres, so it
is worthwhile to assess the distribution of their atoms
across different ionization states.
In Figure~\ref{sahaplot} we illustrate these distributions
for 15~elements of interest.
The temperatures shown cover the range where typical
weak lines on the linear part of the curve-of-growth are formed
in the atmosphere of \hd.
Each element is found in at most three ionization stages.
Figure~\ref{sahaplot} shows that
first ions are the majority for 
most of these elements over the relevant temperature range.
Substantial amounts of neutral 
copper, zinc, arsenic, selenium, cadmium, tellurium,
osmium, platinum, and mercury are also found.
Ionization corrections for the neutral or singly-ionized states of
these atoms should be reliable. 

Only a few percent of the atoms of iron, strontium, molybdenum, and lead 
are found in the neutral state, so abundances derived from these 
neutral species should be viewed with more caution.
Overionization of Fe~\textsc{i} is a well-known phenomenon.
The abundances of molybdenum derived from Mo~\textsc{i} and 
Mo~\textsc{ii} are in good agreement in \hd, albeit from only
one line of Mo~\textsc{i}, but this agreement is encouraging.
\citet{mashonkina12} have performed non-LTE line formation
calculations for Pb~\textsc{i} lines, finding positive 
corrections of a factor of 2 or more.
This underscores the importance of deriving abundances
from the majority species, although 
Pb~\textsc{ii} lines have not been detected in late-type
stellar atmospheres.

The elements shown in Figure~\ref{sahaplot}
are also representative of other heavy elements
with similar ionization potentials and electronic level structures.
For example, europium and lanthanum have similar first and second ionization 
potentials.
They also have many low-lying
levels that contribute similarly to the partition functions,
so they should respond similarly to conditions in the stellar atmosphere.
Other pairs of elements include palladium and tellurium,
barium and strontium, and iridium and platinum.

In general, we are deriving abundances from the species in the 
majority ionization state or from minority species with a 
substantial presence.  
This certainly does not demonstrate that an LTE analysis is
appropriate (e.g., \citealt{mashonkina01}), and
we have not addressed the issue of departures 
from excitation equilibrium.
Nevertheless, this offers some assurance that uncertainties in the 
ionization equilibrium are at least minimal.

\section{Results}
\label{results}

\subsection{UV Absorption Lines}
\label{lines}

\begin{figure}
\begin{center}
\includegraphics[angle=0,width=3.25in]{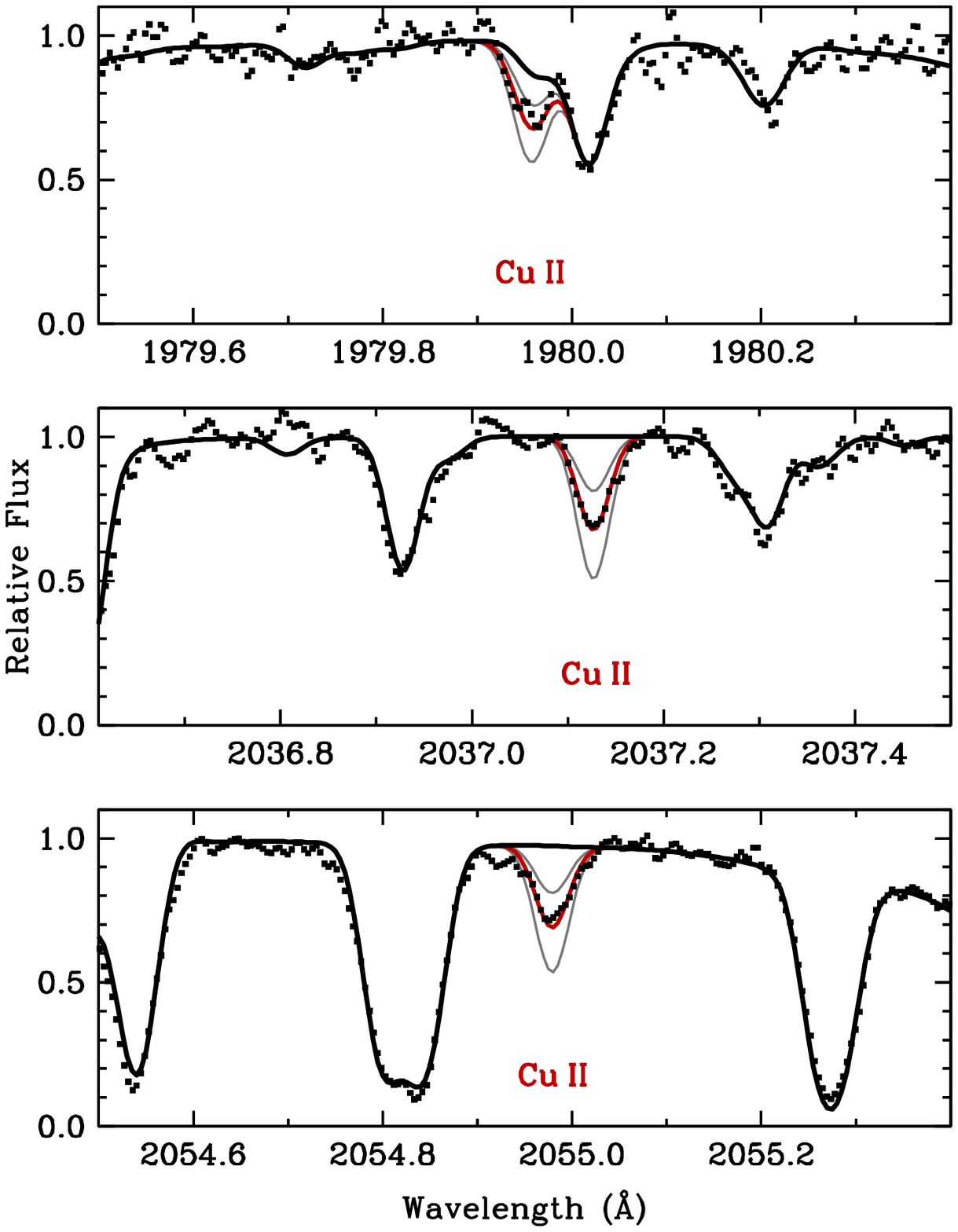}
\end{center}
\caption{
\label{cu1plot}
Synthesis of three Cu~\textsc{ii} lines in the UV.
The observed STIS spectrum is marked by the filled squares.
The best-fit synthesis is indicated by the bold red line,
and the light gray lines show variations in the best-fit abundance
by $\pm$~0.3~dex.
The bold black line indicates a synthesis with no Cu~\textsc{ii} present.
}
\end{figure}

\begin{figure}
\begin{center}
\includegraphics[angle=0,width=3.25in]{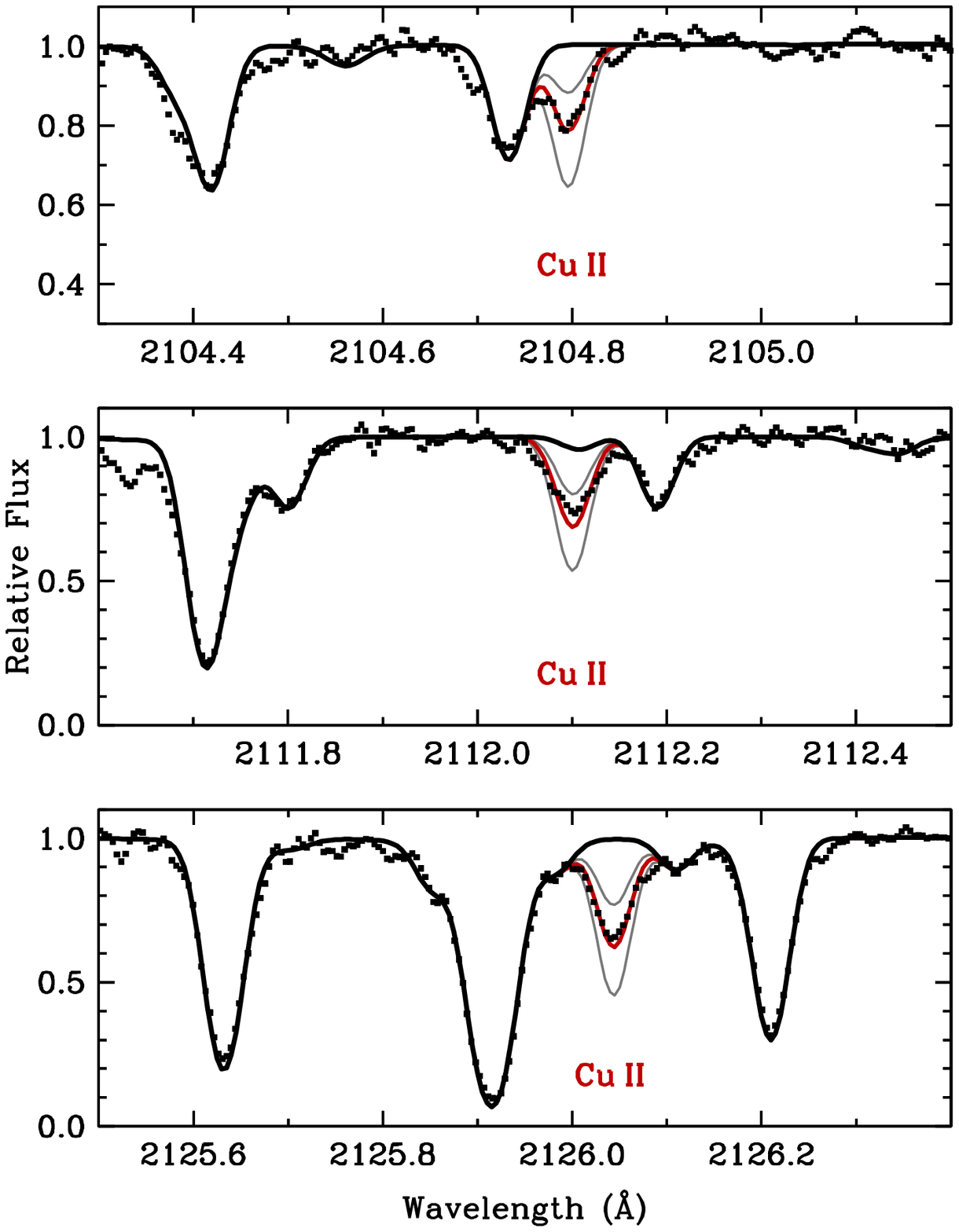}
\end{center}
\caption{
\label{cu2plot}
Synthesis of three Cu~\textsc{ii} lines in the UV.
The observed STIS spectrum is marked by the filled squares.
The best-fit synthesis is indicated by the bold red line,
and the light gray lines show variations in the best-fit abundance
by $\pm$~0.3~dex.
The bold black line indicates a synthesis with no Cu~\textsc{ii} present.
}
\end{figure}

\begin{figure}
\begin{center}
\includegraphics[angle=0,width=3.25in]{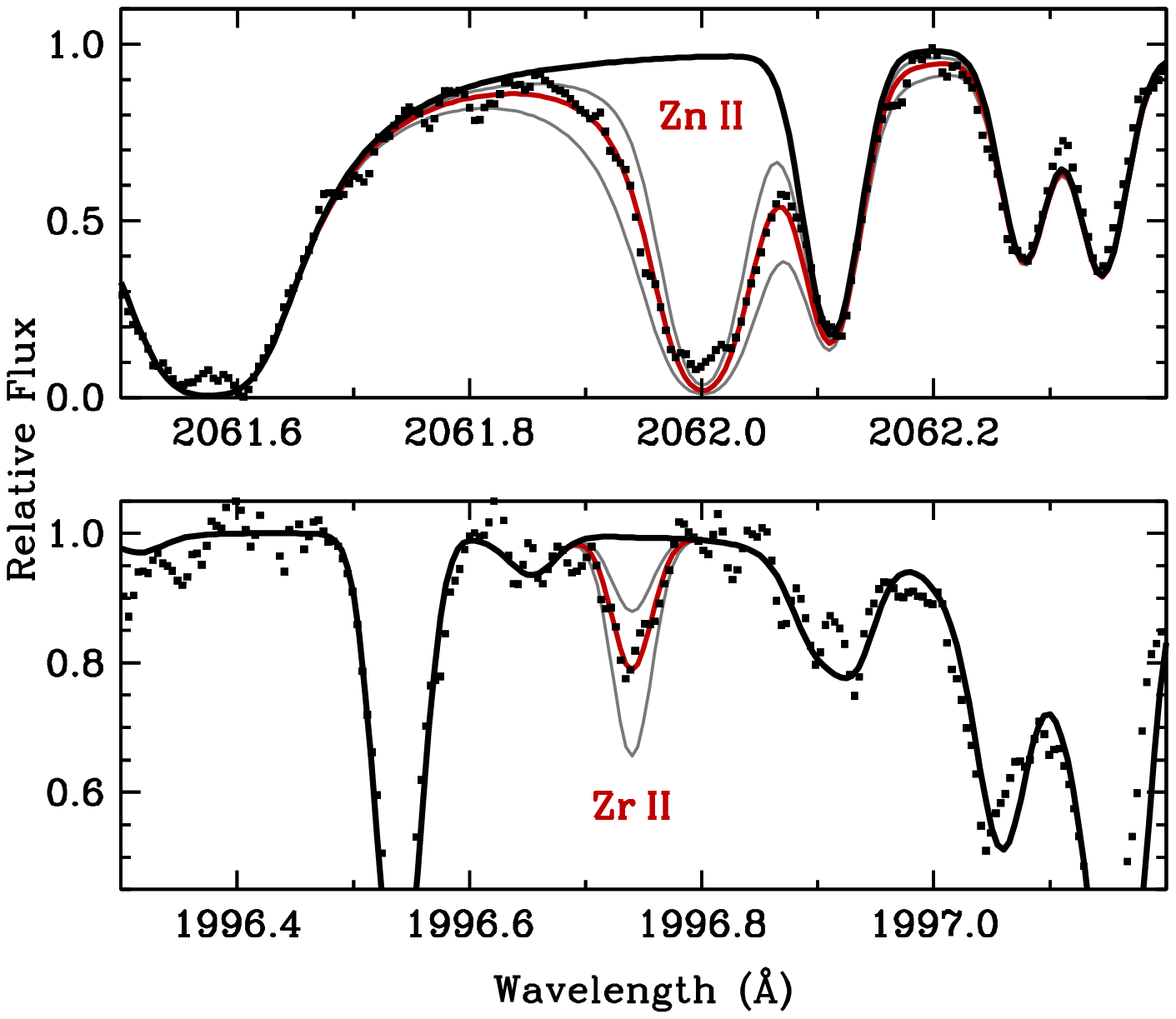}
\end{center}
\caption{
\label{znplot}
Synthesis of the Zn~\textsc{ii} and Zr~\textsc{ii} lines in the UV.
The observed STIS spectrum is marked by the filled squares.
The best-fit synthesis is indicated by the bold red line,
and the light gray lines show variations in the best-fit abundance
by $\pm$~0.3~dex.
The bold black line indicates a synthesis with no Zn~\textsc{ii} or 
Zr~\textsc{ii} present.
}
\end{figure}

\begin{figure}
\begin{center}
\includegraphics[angle=0,width=3.25in]{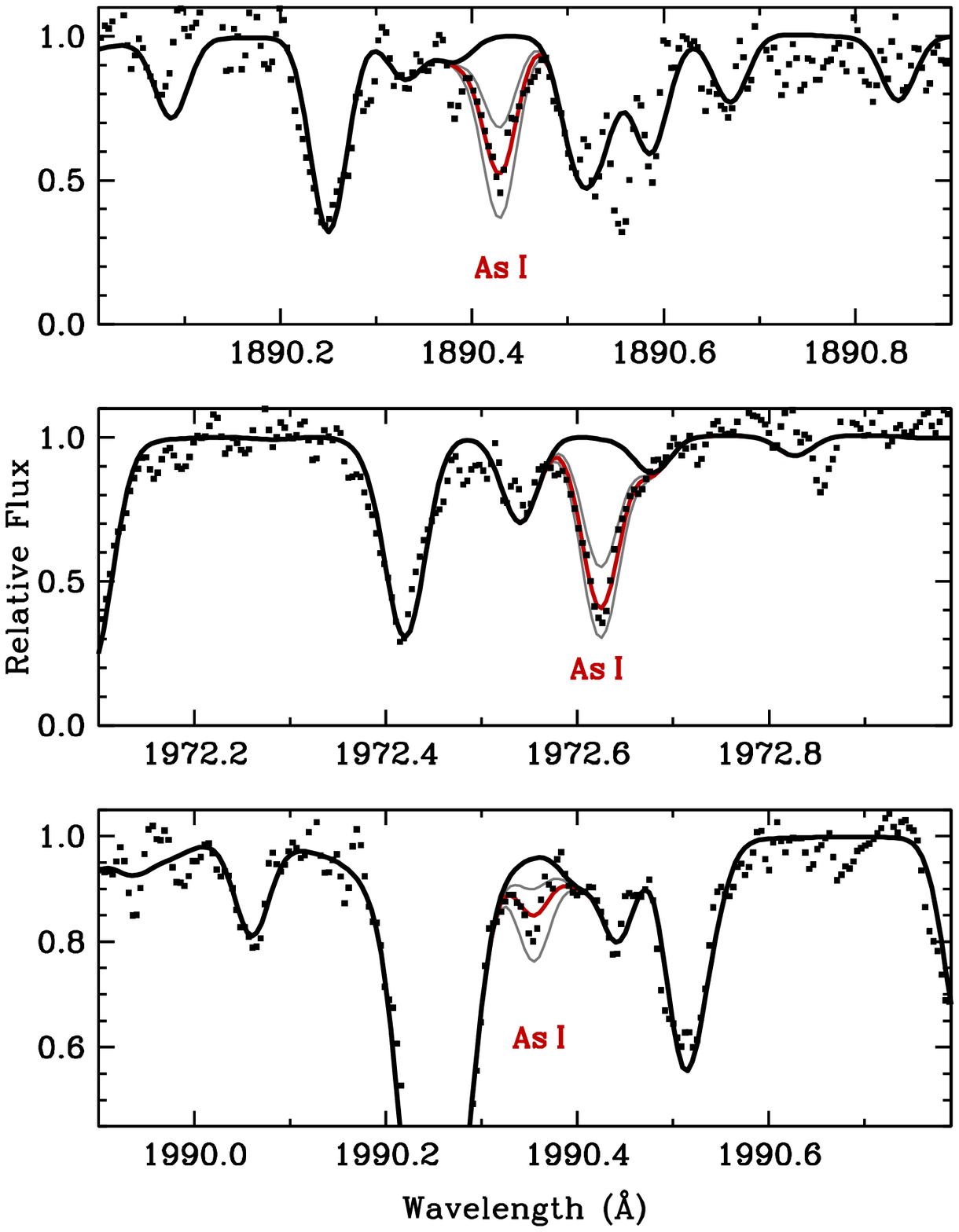}
\end{center}
\caption{
\label{asplot}
Synthesis of the As~\textsc{i} lines in the UV.
The observed STIS spectrum is marked by the filled squares.
The best-fit synthesis is indicated by the bold red line,
and the light gray lines show variations in the best-fit abundance
by $\pm$~0.3~dex.
The bold black line indicates a synthesis with no As~\textsc{i}
present.
}
\end{figure}

\begin{figure}
\begin{center}
\includegraphics[angle=0,width=3.25in]{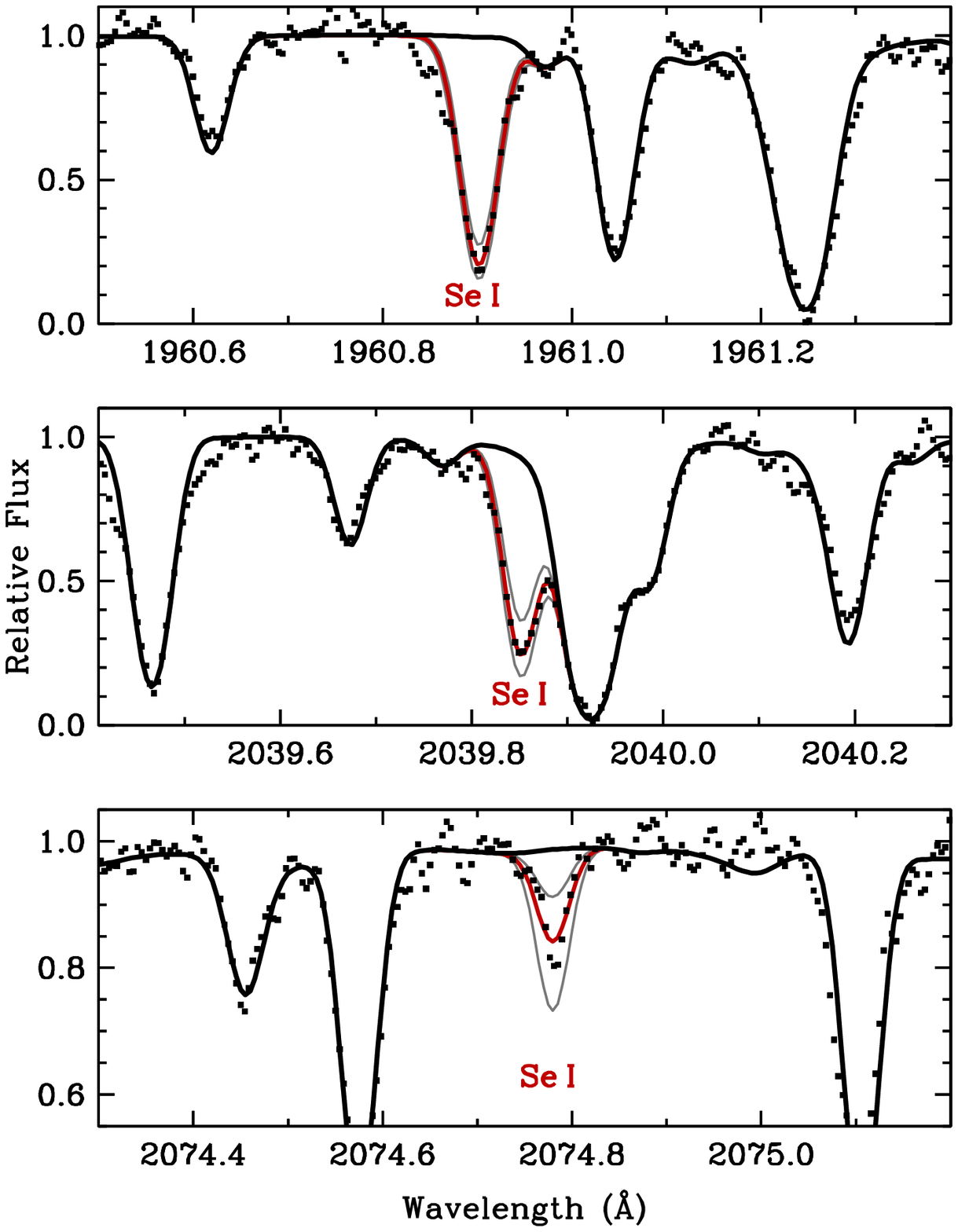}
\end{center}
\caption{
\label{seplot}
Synthesis of the Se~\textsc{i} lines in the UV.
The observed STIS spectrum is marked by the filled squares.
The best-fit synthesis is indicated by the bold red line,
and the light gray lines show variations in the best-fit abundance
by $\pm$~0.3~dex.
The bold black line indicates a synthesis with no Se~\textsc{i}
present.
}
\end{figure}

\begin{figure}
\begin{center}
\includegraphics[angle=0,width=3.25in]{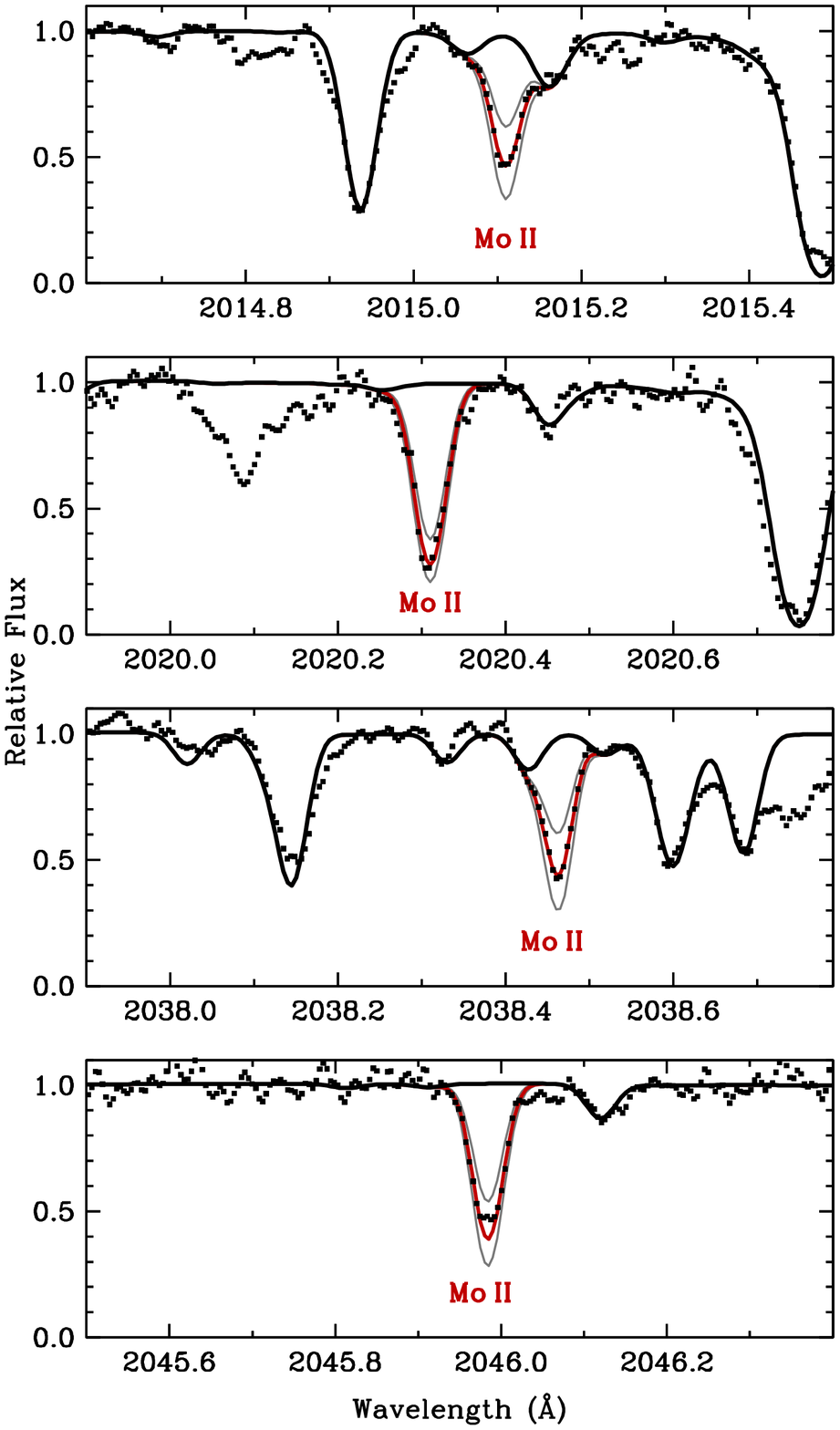}
\end{center}
\caption{
\label{moplot}
Synthesis of the Mo~\textsc{ii} lines in the UV.
The observed STIS spectrum is marked by the filled squares.
The best-fit synthesis is indicated by the bold red line,
and the light gray lines show variations in the best-fit
abundance by $\pm$~0.3~dex.
The bold black line indicates a synthesis with no Mo~\textsc{ii}
present.
}
\end{figure}

\begin{figure}
\begin{center}
\includegraphics[angle=0,width=3.25in]{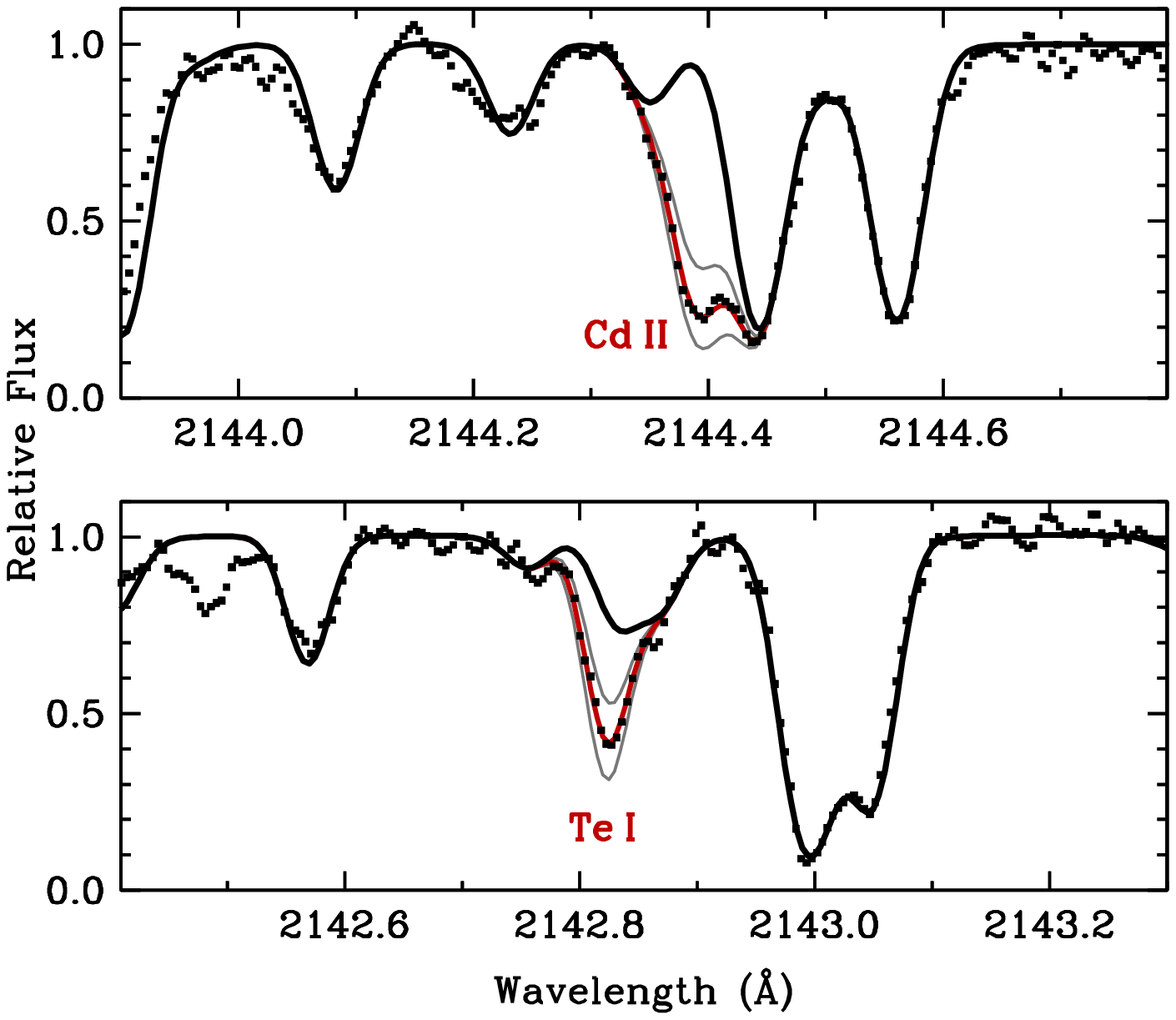}
\end{center}
\caption{
\label{cdplot}
Synthesis of the Cd~\textsc{ii} and Te~\textsc{i} lines in the UV.
The observed STIS spectrum is marked by the filled squares.
The best-fit synthesis is indicated by the bold red line,
and the light gray lines show variations in the best-fit abundance
by $\pm$~0.3~dex.
The bold black line indicates a synthesis with no Cd~\textsc{ii}
or Te~\textsc{i} present.
}
\end{figure}

\begin{figure}
\begin{center}
\includegraphics[angle=0,width=3.25in]{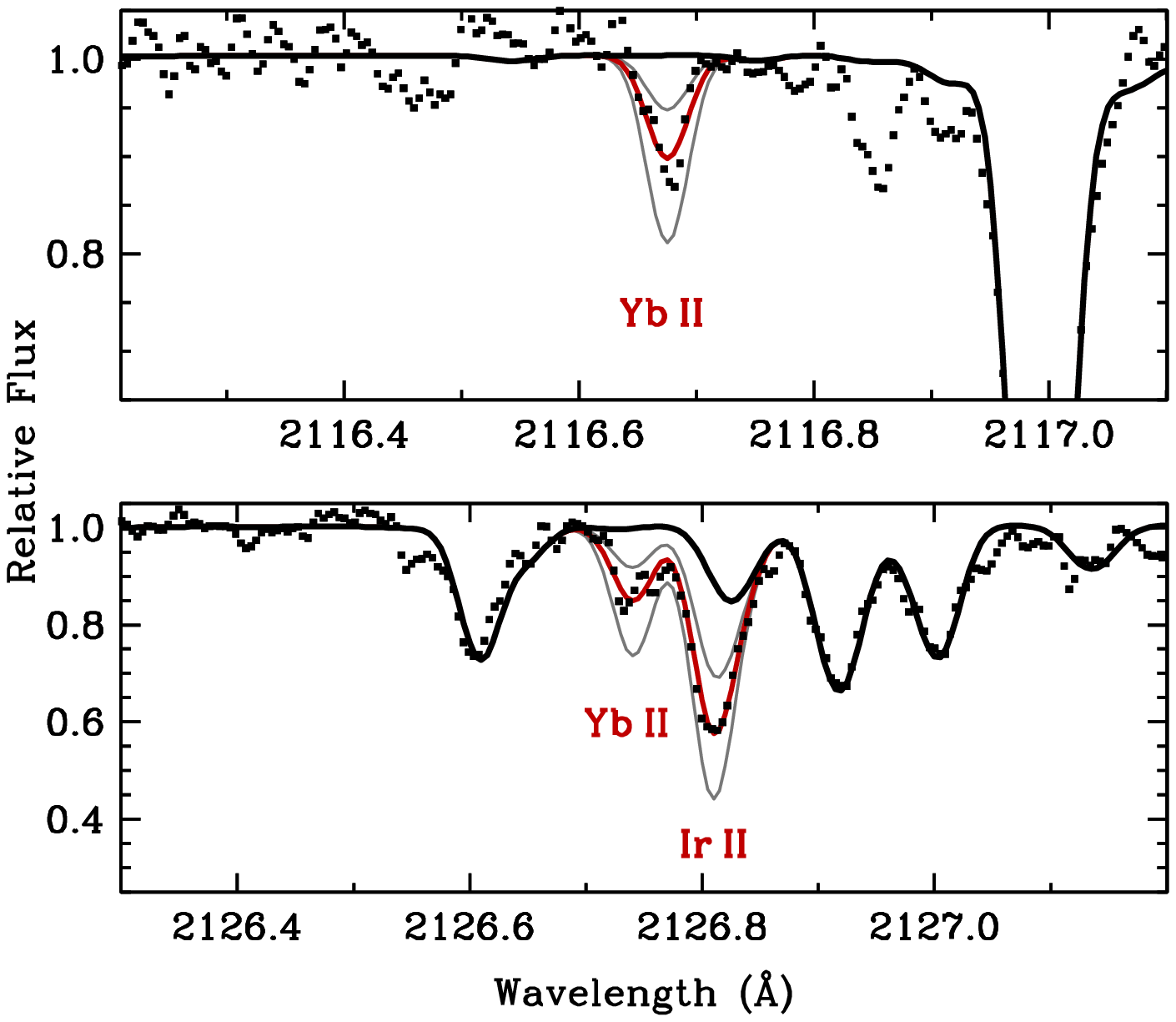}
\end{center}
\caption{
\label{ybplot}
Synthesis of the Yb~\textsc{ii} and Ir~\textsc{ii} lines in the UV.
The observed STIS spectrum is marked by the filled squares.
The best-fit synthesis is indicated by the bold red line,
and the light gray lines show variations in the best-fit abundance
by $\pm$~0.3~dex.
The bold black line indicates a synthesis with no Yb~\textsc{ii}
or Ir~\textsc{ii} present.
}
\end{figure}

\begin{figure}
\begin{center}
\includegraphics[angle=0,width=3.25in]{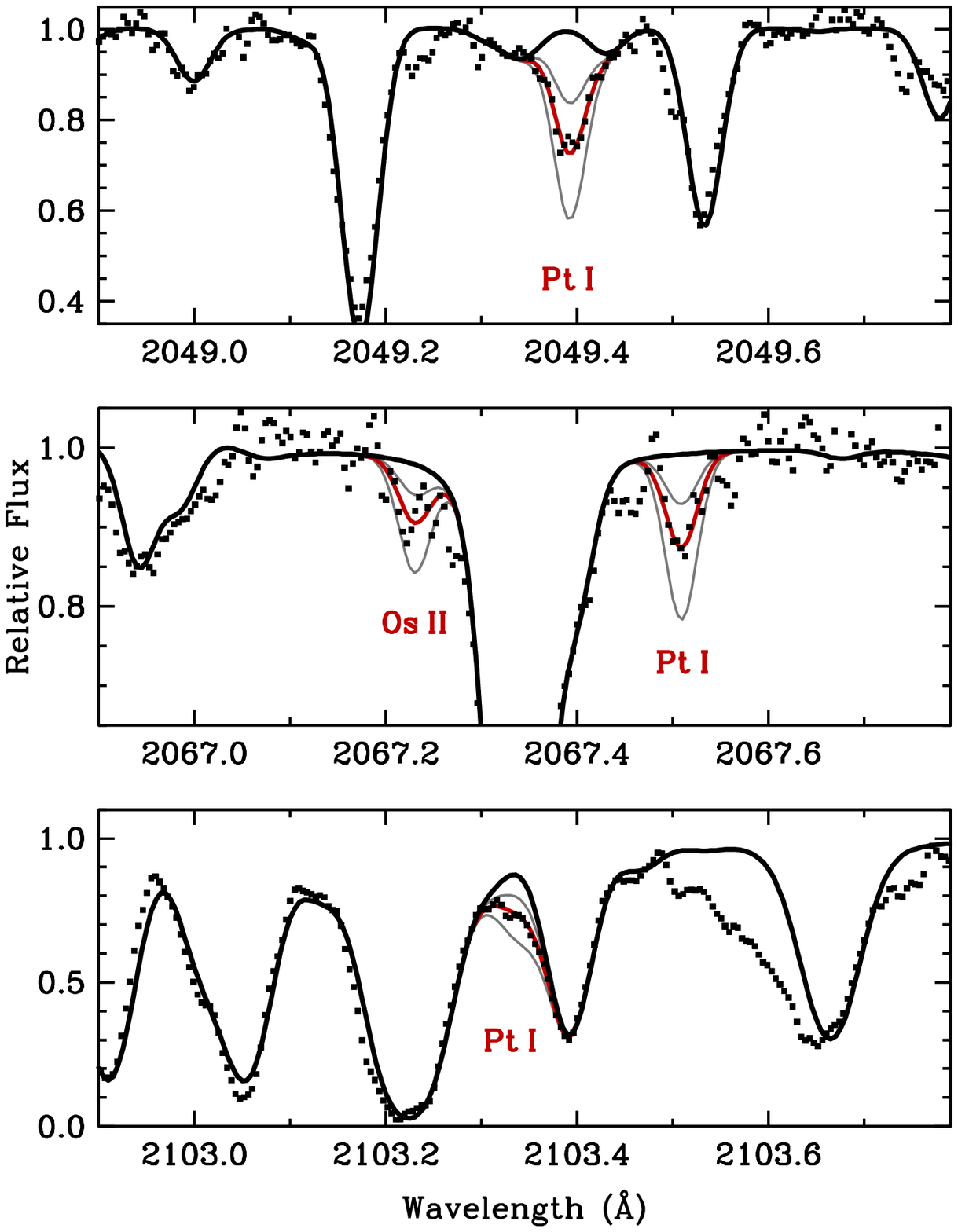}
\end{center}
\caption{
\label{osplot}
Synthesis of the Os~\textsc{ii} and Pt~\textsc{i} lines in the UV.
The observed STIS spectrum is marked by the filled squares.
The best-fit synthesis is indicated by the bold red line,
and the light gray lines show variations in the best-fit abundance
by $\pm$~0.3~dex.
The bold black line indicates a synthesis with no Os~\textsc{ii}
or Pt~\textsc{i} present.
}
\end{figure}

Regions of the STIS spectrum and synthetic spectra fits around the
UV lines of interest are shown in 
Figures~\ref{cu1plot}--\ref{osplot}.
Many of these features are rarely used for stellar abundance analysis.

Four of the six Cu~\textsc{ii} lines detected in the STIS spectrum
of \hd\ are relatively unblended, as shown in 
Figures~\ref{cu1plot} and \ref{cu2plot}.
The linelists of \citet{kurucz95} suggest that
the Cu~\textsc{ii} 1979.956\,\AA\ line might be blended with an
Fe~\textsc{ii} line at 1979.963\,\AA.
No experimental $\log(gf)$ value is available for this line.
The \citeauthor{kurucz95} lists also indicate that
the Cu~\textsc{ii} 2112.100\,\AA\ line might be blended with a weak
Ni~\textsc{ii} line at 2112.107\,\AA.
No experimental $\log(gf)$ value is available for this line, either.
In both cases, we vary the strength of the blending feature
within the limits allowed by the observed line profile.
This introduces considerable uncertainty (0.2--0.3~dex)
into the derived Cu~\textsc{ii} abundance, so we weight
these lines less in the mean copper abundance.
The Cu~\textsc{ii} 2135.98\,\AA\ line is clearly detected in \hd,
but it is strong and blended with another strong Fe~\textsc{i}
line at 2135.96\,\AA.
We do not use this line in our analysis.
The mean abundance derived from the two Cu~\textsc{i} 
resonance lines at 3247\,\AA\
and 3273\,\AA, [Cu/Fe]~$= -$0.87~$\pm$~0.07 (internal), 
is about a factor of 2 lower than
the mean abundance derived from the six Cu~\textsc{ii} UV lines,
[Cu/Fe]~$= -$0.58~$\pm$~0.05 (internal).
Cu~\textsc{ii} is by far the dominant ionization state in the
atmosphere of \hd, and this offset may indicate the 
presence of significant non-LTE effects affecting Cu~\textsc{i}.  

The Zn~\textsc{ii} 2062\,\AA\ line is strong and clearly detected in our
spectrum of \hd, as shown in Figure~\ref{znplot}.
(The 2025.48\,\AA\ Zn~\textsc{ii} line is too blended 
with a strong Mg~\textsc{i} line at 2025.82\,\AA\ 
to be useful as an abundance indicator.)
The wings of the 2062\,\AA\ line offer the most abundance sensitivity, and
we ignore the line core in the fit.
Calculations of the van der Waals damping parameters
for this transition are not available, 
so we resort to the standard \citet{unsold55} approximation.
The zinc abundance derived from this Zn~\textsc{ii} line is in 
excellent agreement with that derived from 6~Zn~\textsc{i} lines
in the optical spectral region.

We detect 3~As~\textsc{i} lines in \hd, as shown in Figure~\ref{asplot}.
All are relatively unblended in \hd.
These lines give somewhat discrepant abundance results
($\sigma =$~0.37~dex).
The experimental transition probability uncertainties are comparable,
and the theoretical values are in good agreement, so
it is not obvious that one of these lines should be preferred
over the others.
We calculate the arsenic abundance in \hd\ by taking
an unweighted mean of these three lines.
The fourth As~\textsc{i} line listed in Table~\ref{astab}, 1937\,\AA, 
is close to the Al~\textsc{i} auto-ionization line at
1936.46\,\AA.
This Al~\textsc{i} line increases the continuous opacity locally 
and is not well modeled by our approach, so
we do not use the As~\textsc{i} 1937\,\AA\ line as an abundance indicator.

We detect 3~Se~\textsc{i} lines in \hd, as shown in Figure~\ref{seplot}.
The 1960\,\AA\ line is clearly saturated and not 
on the linear part of the curve of growth.
The 2039\,\AA\ line is blended with a Cr~\textsc{i} line at 2039.92\,\AA,
which can be fit by our synthesis.
The 1960\,\AA\ and 2074\,\AA\ lines are unblended in \hd.
All 3~lines give similar abundance results ($\sigma =$~0.08).

We detect one unblended line of Zr~\textsc{ii} in the STIS spectrum of \hd,
1996\,\AA, as shown in Figure~\ref{znplot}.
The abundance derived from this line is in excellent agreement with
the abundance derived from the 19~Zr~\textsc{ii} lines in the
optical spectral range (3095\,\AA--4208\,\AA).
This indicates that any possible offset between the optical and UV
abundance scales (due to, e.g., inaccurate calculation of the 
continuous opacity) is small.

We detect 4~Mo~\textsc{ii} lines in \hd, as shown in Figure~\ref{moplot}.
These lines are relatively clean and provide similar abundance
results ($\sigma =$~0.09).
The abundance derived from these lines agrees with the abundance derived
from the Mo~\textsc{i} line at 3864\,\AA, although we recall 
from Section~\ref{saha} that very little neutral molybdenum is 
present in the line-forming layers of \hd.
\citet{peterson11} also derived a molybdenum abundance from these
Mo~\textsc{ii} lines in the same STIS spectrum of \hd.
After correcting for the different $\log(gf)$ scales used, our
[Mo/Fe] ratio is about 0.4~dex lower than hers.
This discrepancy is larger than can be accounted for by the slight 
differences in model atmosphere parameters
($\delta T_{\rm eff} =$~50~K, 
 $\delta \log g =$~0.1,
 $\delta v_{t} =$~0.1~\kmsec),
and the line-by-line differences cannot be assessed from the
published data.
Regardless, the molybdenum in \hd\ is still overabundant,
a curious fact that \citeauthor{peterson11} discussed in detail.
The possibility that a small amount of \spro\
material may be present in \hd\ (Section~\ref{heavy})
is an unlikely explanation for the molybdenum overabundance.

We detect the Cd~\textsc{ii} 2144.39\,\AA\ line in \hd, as shown in
Figure~\ref{cdplot}.
This line is blended with Fe~\textsc{ii} 2144.35\,\AA\ 
and Fe~\textsc{i} 2144.44\,\AA.
These blends can be adjusted to fit the line profile.
The residual absorption can be well fit by the Cd~\textsc{ii} line.

The Te~\textsc{i} 2142.822\,\AA\ line in \hd\ is very blended with
Fe~\textsc{i} 2142.832\,\AA.
No experimental $\log(gf)$ value is known for the Fe~\textsc{i} line.
These lines are separated just enough to 
break the $\log(gf)$-abundance degeneracy in the high-resolution
STIS spectrum.
We estimate the strength of the Fe~\textsc{i} line by
fitting the line profile with a combination of Te~\textsc{i} and 
Fe~\textsc{i} absorption.
We include
absorption from an extra, unidentified line at 2142.87\,\AA.
We can produce a reasonable fit to the Te~\textsc{i} 2142\,\AA\
line, but we caution that the uncertainty in the derived
abundance surely is larger than the uncertainties of other,
unblended absorption features.

We detect 2~Yb~\textsc{ii} absorption lines in the STIS spectrum, 
2116\,\AA\ and 2126\,\AA, 
shown in Figure~\ref{ybplot}, as well as the more common
optical Yb~\textsc{ii} line at 3694\,\AA.
The abundance derived from the 2116\,\AA\ line is a factor of 2 
greater than that derived from the other two lines,
which are in good agreement.
The general agreement is reassuring for studies where only
the 3694\,\AA\ line is available.

We report a marginal detection of the Os~\textsc{ii} line at
2067\,\AA, as shown in Figure~\ref{osplot}.
This line is weak and the spectrum is noisy here,
so the abundance should be interpreted with caution.
The other Os~\textsc{i} or Os~\textsc{ii} abundance indicators 
commonly detected in \rpro\ rich metal-poor stars
are not covered by existing spectra of \hd.

We detect one Ir~\textsc{ii} line in \hd, as shown in Figure~\ref{ybplot}.
This line appears to be blended with a weak, high-excitation 
Ni~\textsc{ii} line at 2126.83\,\AA.
No experimental $\log(gf)$ value is available for this Ni~\textsc{ii} line.
We fit the line profile as best as possible.
None of the common UV Ir~\textsc{i} abundance indicators
are covered by existing spectra of \hd.

We detect 3~Pt~\textsc{i} lines in \hd, as shown in Figure~\ref{osplot}.
Two of these lines, 2067\,\AA\ and 2049\,\AA, 
are relatively clean.
The third line, 2103\,\AA, is detected as extra absorption between two
stronger lines.  
Our syntheses do not predict absorption from other species between
2103.5\,\AA\ and 2103.65\,\AA, whereas the observed spectrum
clearly indicates absorption.
The abundance derived from this Pt~\textsc{i} line is much less secure,
and we weight it half as much as the other two lines in the
average.
Nevertheless, the abundances of all three lines are in good agreement
($\sigma =$~0.10).

We do not detect the Hg~\textsc{ii} 1942.27\,\AA\ absorption line
in \hd.
We synthesize the line and derive an upper limit on the mercury abundance.

\subsection{Other Abundances}

We have derived abundances for all Fe-group elements from calcium to zinc.
We have detected multiple ionization stages for several of these elements,
including titanium, vanadium, chromium, manganese, iron, copper, and zinc.
The [Ti/Fe], [V/Fe], [Mn/Fe], and [Zn/Fe] ratios
derived from neutral atoms and single ions are in agreement in \hd.
These elements are mostly ionized in the line-forming regions of \hd,
so this agreement is gratifying.
We recover the persistent disagreement 
between [Cr/Fe] ratios derived from neutral atoms and ions 
(e.g., \citealt{gratton91,sobeck07,bergemann10}, and references therein),
[Cr~\textsc{ii}/Fe~\textsc{ii}]$-$[Cr~\textsc{i}/Fe~\textsc{i}]
$= +$0.31~dex.
We also find a similar offset between neutral and ionized copper,
[Cu~\textsc{ii}/Fe~\textsc{ii}]$-$[Cu~\textsc{i}/Fe~\textsc{i}]
$= +$0.29~dex.

\citet{hansen11} studied the palladium and silver abundances in a
large sample of dwarf and giant metal-poor stars, including \hd.
Our [Pd/Fe] ratio is in good agreement with theirs.
They only reported an upper limit on [Ag/Fe].
We report
a tentative detection of the weak Ag~\textsc{i} line at 3382\,\AA.
The abundance derived from this line is consistent with their 
reported upper limit.
Most of the silver in the line-forming
layers of \hd\ is Ag~\textsc{ii}, so our abundance should be
viewed with caution.
The lower-lying levels of Ag~\textsc{i} are all known, fortunately,
so the partition functions of Ag~\textsc{i} should be relatively complete.
The first ionization potential of silver is 7.58~eV, nearly the same as
copper, 7.73~eV, which has a similar electronic configuration.
By analogy, 
Figure~\ref{sahaplot} suggests that no more than a few percent of
the silver is present as Ag~\textsc{i}.
The strongest transitions of Ag~\textsc{ii} lie deep in the vacuum UV, and
there is little hope of detecting these lines in late-type stars.
The first ionization potential of palladium is somewhat higher, 8.34~eV,
but not as high as cadmium, 8.99~eV,
so a small amount of Pd~\textsc{i} may be present.

\section{Discussion}
\label{discussion}

\subsection{Nucleosynthesis of Arsenic and Selenium and Their
Abundances beyond the Solar System}

CI chondrite meteorites reveal that selenium 
is the second most abundant element heavier than zinc in the S.S.;
arsenic is the ninth most abundant.
Isotopes of each can be produced by both \spro\ and \rpro\
nucleosynthesis.
In S.S.\ material, the \spro\ contribution to these elements
is predicted to be small, $\approx$~5--10\% 
(e.g., \citealt{arlandini99,bisterzo11}).
While the flow of \rpro\ nucleosynthesis should pass through
the parent isobars of these elements,
other mechanisms may also contribute to their production in
environments traditionally ascribed to the \rpro.
These include
a charged-particle ($\alpha$) process
(e.g., \citealt{woosley92,farouqi09})
or a weak \rpro\ in
nuclear statistical equilibrium with a small neutron excess
\citep{wanajo11}.
Calculations of \rpro\ nucleosynthesis using the ``waiting-point''
approximation and a weighting of neutron density components
can reproduce the predicted S.S.\ \rpro\ elemental abundance
of selenium
(e.g., \citealt{kratz07,farouqi09}).

The elements arsenic and selenium have not been detected 
previously in halo stars, and neither has been detected in the 
solar photosphere.
As~\textsc{i} and \textsc{ii} lines in the UV have been detected in 
chemically-peculiar stars, 
including $\chi$ Lupi (e.g., \citealt{leckrone99}, and references therein).
\citet{cardelli93} detected As~\textsc{ii} and Se~\textsc{ii} in
the interstellar medium (ISM) towards $\zeta$~Oph.
[Se~\textsc{iv}] is detected in emission in planetary nebulae, 
where its abundance can 
frequently be attributed to self-enrichment by the \spro\
\citep{dinerstein01,sterling08}.
This same line has also been found
in ultra-compact H~\textsc{ii} regions in the vicinity of hot O stars
\citep{blum08,romanlopes09}.
\citet{chayer05} reported the detection of several high-ionization 
absorption lines of arsenic and selenium in the far UV spectrum 
of a hydrogen-deficient white dwarf, 
although \citet{boyce08} suggest that these are likely 
misidentifications of ISM lines or photospheric
lines of lighter ions.
None of these detections offers the same opportunity that \hd\ does
to study detailed (\rpro) nucleosynthesis patterns including
arsenic and selenium.

\subsection{How Normal Is HD 160617?}

\hd\ is an old (11.7~$\pm$~2.1~Gyr) star with kinematics consistent
with membership in the Galactic halo \citep{holmberg07}.
\citeauthor{holmberg07}\ also find that \hd\ 
shows no evidence of radial velocity variations
over a long (12~yr) baseline,
indicating that it is not in a binary system.
\citet{demedeiros06} find a low rotational velocity 
($v \sin i =$~6.2~$\pm$~2.6~\kmsec)
for \hd.
These are typical characteristics for field stars at 
the metallicity of \hd, [Fe/H]~$= -$1.8.
Several studies of the composition of \hd\ over the last 30~years
have suggested that abundances of 
some of the light ($Z \leq$~30) elements in this star 
may be anomalous.
Here, we reconsider these matters in light of the
high-quality spectra of this star now available.

The $\alpha$ and Fe-group abundances in \hd\ 
fall well within the usual trends for stars at this metallicity,
as established by previous studies
\citep{gratton91,fulbright02,sobeck06,nissen07,primas08,roederer10b}.
The [O/Fe], [Mg/Fe], [Si/Fe], [Ca/Fe], and [Ti/Fe] ratios 
are all enhanced by factors
of 1.6 to 2.8, typical for metal-poor halo stars enriched by 
core-collapse supernovae.
\citet{nissen07} found that the [S/Fe] ratio in \hd\ is also 
enhanced by a similar factor.
The [Na/Fe] ratio in \hd\
lies on the upper envelope of the [Na/Fe] ratios found in field stars,
but there is considerable scatter in [Na/Fe] at this metallicity
(e.g., \citealt{pilachowski96,gratton00}).
The abundance ratios among Fe-group elements are unremarkable.

Observations by \citet{bessell82} and \citet{laird85} found 
[N/Fe] to be significantly enhanced in \hd\ by a factor of 50--100
relative to the solar ratio.
Subsequent studies by \citet{zhao90} and \citet{johnson07} have
found much lower levels of nitrogen enhancement (factors of 2.5--6).
Our nitrogen abundance, derived from NH, supports the lower value.
Nevertheless, this [N/Fe] ratio is still somewhat higher than that
found in typical field stars
(e.g., \citealt{gratton00,israelian04}).
The [C/Fe] ratio in \hd, derived by us from CH, is approximately solar.
We are limited by systematic uncertainties 
inherent to deriving abundances from
hydride bands (continuum placement or 
large corrections from a 1-dimensional
to a 3-dimensional model atmosphere, for example)
and offsets introduced by using different abundance indicators
(e.g., \citealt{tomkin92}).
We shall not dwell on the issue of carbon and nitrogen abundances 
except to say that nitrogen may be mildly enhanced in \hd.

\citet{primas98} found that the boron abundance in \hd\ is 
unusually low for its evolutionary state, yet the lithium
abundance in \hd\ is consistent with that of the Spite plateau
for metal-poor dwarf and subgiant stars.
Those authors noted that the abundance of beryllium, 
the element between lithium and boron, in \hd\ is
uncertain. 
Examination of Figure~3 of \citet{molaro97} reveals that 
the S/N of the UVES spectrum employed in the present study
is significantly higher than what was available previously.
Using the UVES spectrum, we derive a beryllium abundance
from the Be~\textsc{ii} 3130.42\,\AA\ and 3131.07\,\AA\ doublet.
The hfs of these lines can be neglected in the abundance analysis.
This abundance is in good agreement with more recent work
\citep{tan09} and falls well within the Galactic trends of 
$A$(Be) versus [Fe/H] or [O/H] (e.g., \citealt{boesgaard11}).
This suggests that beryllium is not depleted in \hd\
relative to other halo stars.

The unexplained low boron abundance and slight nitrogen enhancement
notwithstanding, 
the kinematics and composition of \hd\ suggest that it is
a normal, metal-poor halo star passing through the Solar neighborhood.

\subsection{The Heavy Element Abundance Pattern in \hd}
\label{heavy}

\begin{figure*}
\begin{center}
\includegraphics[angle=270,width=4.5in]{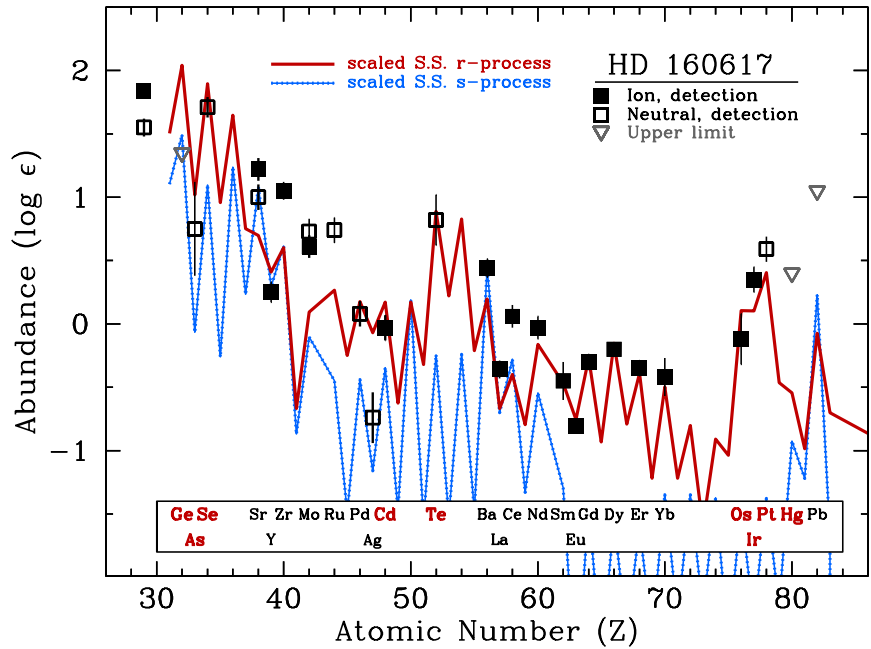}
\end{center}
\caption{
\label{logepsplot}
Logarithmic abundances in \hd\ as a function of atomic number.
Abundances derived from neutral species are shown as open squares,
and abundances derived from singly-ionized species are shown
as filled squares.
Error bars represent internal uncertainties only.
The abundances of neutral species have been offset vertically in this
figure by $+$0.15~dex to reflect the difference between [Fe~\textsc{i}/H]
and [Fe~\textsc{ii}/H].
Open, downward-facing triangles mark upper limits.
The bold, solid red curve indicates the scaled S.S.\ \rpro\ residuals,
scaled to match the heavy rare earth element abundances (Sm--Yb)
in \hd.
The studded blue curve indicates the scaled S.S.\ \spro\ predictions,
scaled to match the barium abundance in \hd.
These predictions are taken from \citet{sneden08}, but the
lead and bismuth predictions have been taken from the low-metallicity
stellar models of \citet{bisterzo11}.
Symbols of elements that can only be detected 
in the UV from space-based spectra
are shown in bold red letters in the box at the bottom.
 }
\end{figure*}

Figure~\ref{logepsplot} shows the heavy element abundance pattern
in \hd.
In the legend, we highlight elements that can only be detected
in the UV using \textit{HST}.
These elements comprise 30\% of all heavy elements detected in \hd,
including all of the elements detected at the three \rpro\ peaks.
This striking illustration demonstrates the unique opportunity
made available by placing sensitive UV echelle spectrographs in space.

In Figure~\ref{logepsplot},
abundances derived from neutral atoms are shown with open squares, and
abundances derived from ions are shown with filled squares.
To account for the difference between [Fe~\textsc{i}/H] and 
[Fe~\textsc{ii}/H] in \hd, for display purposes in Figure~\ref{logepsplot}
we shift the abundances derived from neutral atoms 
upward by 0.15~dex.
The overall normalization between the
neutrals and ions is somewhat uncertain, so ratios between
elements detected in different ionization states should be
interpreted with caution
(e.g., [Se/Zr], [Pd/Cd], [Cd/Te], [Te/Eu], and [Pt/Ir]).
Ratios constructed among elements derived from the same ionization state
should be much more secure
(e.g., [As/Se], [Se/Te], [La/Eu], and [Te/Pt]).

\begin{figure*}
\begin{center}
\includegraphics[angle=270,width=4.5in]{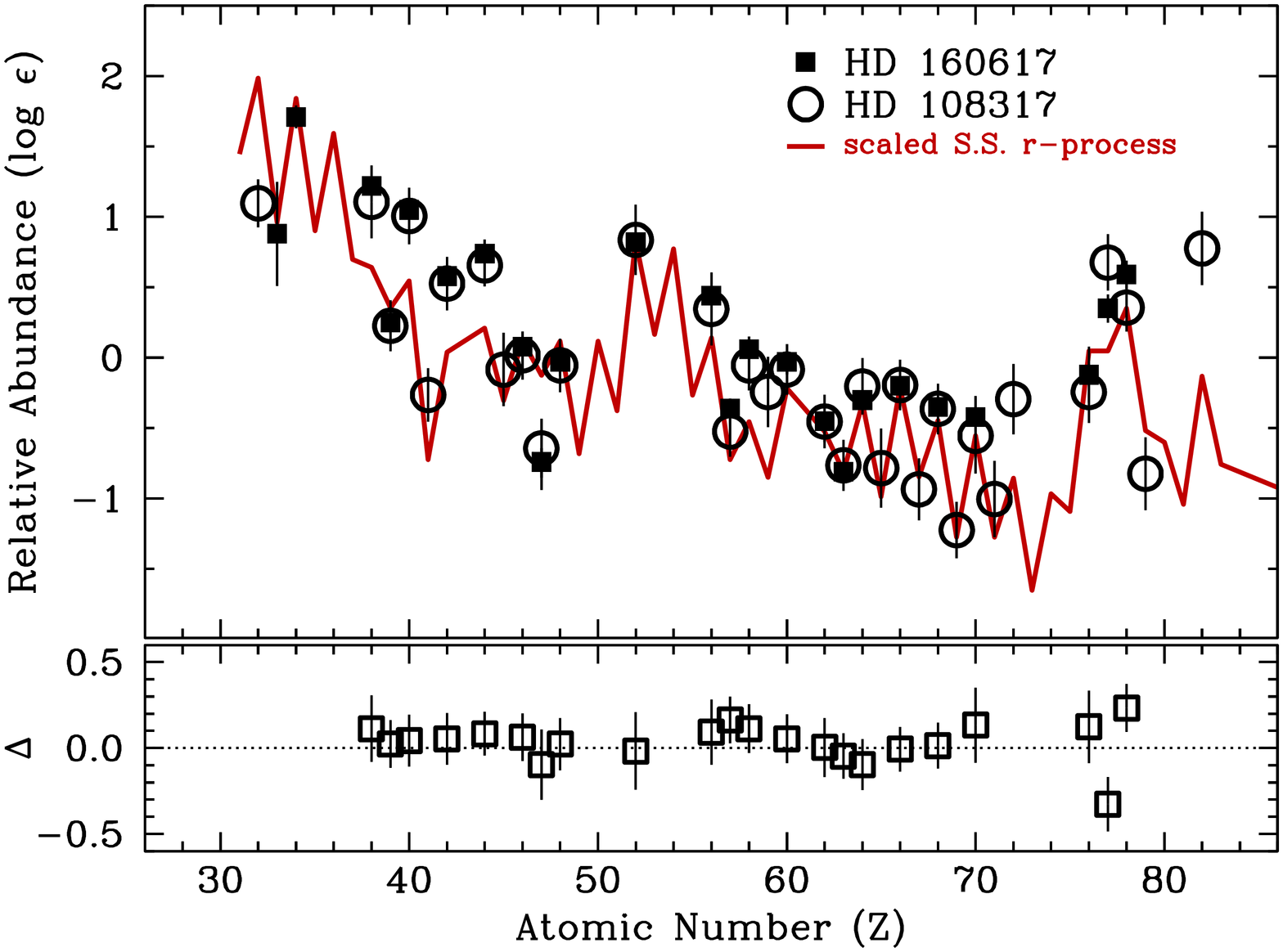}
\end{center}
\caption{
\label{compareplot}
Comparison of the abundances in \hd\ and \hda\ \citep{roederer12b}.
The filled squares indicate the abundances in \hd, the open circles
indicate the abundances in \hda, and the solid red line 
marks the scaled S.S.\ \rpro\ residuals, normalized to the
heavy rare earth abundances (Sm--Yb) in \hd.
The abundances of neutral species in \hd\ and \hda\ 
have been offset vertically in this
figure by $+$0.15~dex and $+$0.18~dex 
to reflect the differences between [Fe~\textsc{i}/H] and [Fe~\textsc{ii}/H].
The open squares in the
bottom panel show the differences between these two stars
(in the sense of \hd\ minus \hda).
Uncertainties in the bottom panel are computed as
(($\sigma_{\rm HD\,160617}^{2} + \sigma_{\rm HD\,108317}^{2}$)/2)$^{1/2}$.
 }
\end{figure*}

Scaled versions of the predicted S.S.\ $s$- and \rpro\ distributions
are also shown in Figure~\ref{logepsplot}.
There are natural limitations to the utility of these predictions, 
but they are useful first approximations to interpret the
heavy element enrichment patterns.
The \rpro\ distribution provides a fair fit to the
observed abundance pattern in \hd, 
while the \spro\ distribution does not.
The \rpro\ distribution matches the elements
at the heavy end of the rare earth domain (samarium through ytterbium) and
the third \rpro\ peak elements (osmium through platinum).
When normalized to these elements, 
the elements at the light end of the rare earth domain
(barium through neodymium)
appear slightly overabundant in \hd.
Some of the lighter elements 
(strontium, zirconium, molybdenum, and ruthenium)
also appear overabundant, others
nearly match the predictions
(arsenic, selenium, yttrium, palladium, cadmium, and tellurium),
and others (germanium, silver)
appear underabundant.
No alternative normalization of the scaled S.S.\ \rpro\ distribution
can provide a satisfactory match to all the elements from
germanium to tellurium.

At least one element at each \rpro\ peak has been detected
in the neutral state: selenium, tellurium, and platinum.
We recall from Section~\ref{lines} that the only line of Te~\textsc{i}
available in our spectrum of \hd\ is quite blended, so the
uncertainty here is considerably larger.
All three of these abundances are close to their predicted
values in the scaled S.S.\ \rpro\ residuals.
If the abundances of osmium and iridium, derived from transitions
among ions, are also considered, the overall agreement
of the three \rpro\ peaks must be considered quite good.

In Figure~\ref{compareplot}, we compare the abundances in \hd\
with those in \hda\ \citep{roederer12b}.
The overall metallicity of \hda\ is a factor of 4 lower than that of \hd,
but both stars have very similar heavy element enrichment levels,
[Eu/Fe]~$\approx +$0.4.
Despite its overall lower metallicity,
we were able to detect more elements in \hda\ because it is
a red giant with a lower continuous opacity.
For the 22~elements in common from strontium to platinum, 
the overall agreement between the two stars is striking.
We previously suggested that \hda\ may 
contain a small amount of \spro\ material in addition to 
a substantial \rpro\ foundation.
The predicted abundances for the
elements at the three \rpro\ peaks do not change
appreciably if a small amount of \spro\ material is present
\citep{roederer12b}.
Another possible explanation is that both patterns are produced by 
\rpro\ nucleosynthesis (or mechanisms that may occur alongside the
\rpro, such as charged-particle freeze-out; 
e.g., \citealt{woosley92,farouqi10}).
\hd\ and \hda\ could be considered 
intermediate cases between the extreme \rpro\ abundance patterns found
in the metal-poor stars \mbox{CS~22892--052} \citep{sneden03}
and \hdb\ \citep{honda06,roederer12b}.

This overall agreement between \hd\ and \hda\ allows us 
to confront systematic effects in the abundance analysis.
In \hd, the cadmium abundance has been derived from one line of Cd~\textsc{ii}.
In \hda, the cadmium abundance has been derived from one line of Cd~\textsc{i}.
The fact that both lines give similar relative abundances
may hint that both indicators are generally reliable.
Similarly, the agreement between the relative tellurium abundances
that have been derived from two different Te~\textsc{i} lines
(2142\,\AA\ and 2385\,\AA) lends confidence to both lines.
For the third \rpro\ peak elements osmium, iridium, and platinum,
the agreement is less good.
Critical comparison of the relative abundance patterns 
among these three elements may not yet be warranted.

Strong odd-even $Z$ effects are observed in both \hd\ and \hda.
These effects are predicted by 
the scaled S.S.\ \rpro\ residuals for the rare earth elements
between the second and third \rpro\ peaks.
These effects are generally not predicted for the elements
from strontium through cadmium.
This phenomenon has noticed before
\citep{sneden00,johnson02,farouqi09,roederer10a},
but the growing number of elements detected from multiple transitions
of the dominant ionization stage lends considerable weight
to its existence.
Stronger odd-even effects are more commonly associated with
\spro\ nucleosynthesis.
The amount of \spro\ material
required to reproduce the observed effect in \hd\ and \hda\ 
exceeds the total abundance of, e.g., strontium, zirconium, or
barium in these stars. 
We regard an \spro\ origin for this odd-even effect as unlikely.

We detect a similar odd-even effect for arsenic and selenium.
This is predicted by the scaled S.S.\ \rpro\ residuals,
though the application of the \rpro\ residual calculation
method at these low atomic numbers is questionable.
The germanium abundance in \hda\ and our upper limit in \hd\
suggest that the odd-even effect 
does not extend to this transition element.
The germanium abundances in these two stars are consistent with
the trends identified by \citet{cowan05}, indicating 
a correlation between [Ge/H] and [Fe/H].
Simulations of nucleosynthesis in the high-entropy wind of
core-collapse supernovae predict that germanium production should
be decoupled from the main \rpro\
(e.g., \citealt{farouqi09}), as is observed.
These simulations predict that arsenic and selenium should also 
be decoupled, which appears to contrast with our observations of \hd.
The fact that arsenic and selenium more-or-less conform to the scaled
S.S.\ \rpro\ residuals, while germanium clearly deviates,
could indicate that this is
the point where the \rpro\ 
turns on.
Additional observations of these three elements in other
metal-poor stars will be necessary to help resolve the matter.

\section{Conclusions}
\label{conclusions}

We have used archive space- and ground-based spectra 
to study detailed abundances
in the metal-poor subgiant star \hd.
We have derived abundances of 51~species of 42~elements in 
\hd, plus upper limits for 3 more.
This inventory includes 27~elements heavier than zinc 
that are produced at least in part by \ncap\ reactions.
For the first time, we have detected elements at all three
\rpro\ peaks in a Galactic halo star:
the first \rpro\ peak element selenium, 
the second \rpro\ peak element tellurium, and 
the third \rpro\ elements osmium, iridium, and platinum. 
This advance is made possible by the high resolution 
UV spectrum of \hd\ taken with STIS on \textit{HST},
covering 1879~$< \lambda <$~2148\,\AA.
We review the literature and present up-to-date transition
probabilities and, when available, line component patterns
including the effects of hfs and IS
for Cu~\textsc{ii}, Zn~\textsc{i} and \textsc{ii},
As~\textsc{i}, Se~\textsc{i}, Mo~\textsc{ii}, 
Cd~\textsc{ii}, Yb~\textsc{ii}, Pt~\textsc{i}, and
Hg~\textsc{ii}.

The heavy elements in \hd\ were produced primarily by \rpro\ 
nucleosynthesis.
When the scaled S.S.\ \rpro\ residuals are normalized to the
rare earth elements in \hd, the elements
at the three \rpro\ peaks have abundances that nearly match
the scaled S.S.\ \rpro\ residuals.
This result reaffirms, on the basis of critical elements
detected for the first time, many previous suggestions that
the nucleosynthesis that occurred in the early universe
resembles later \rpro\ events that produced about half the
heavy elements in the S.S.
On the other hand, this result 
may be somewhat surprising, since the production
of arsenic and selenium is predicted to be decoupled from
that of the \rpro\ elements at the second peak and beyond.
Our non-detection of germanium 
in \hd\ indicates that this element does not follow the scaled S.S.\
\rpro\ residuals.
We leave it as a challenge to theory to find a 
nucleosynthesis model capable of reproducing
the observed abundance patterns in this star.

\acknowledgments

We thank the referee for providing an extremely rapid 
and helpful report, and we thank C.\ Hansen and J.\ Sobeck
for sending results in advance of publication.
I.U.R.\ thanks G.\ Preston for 
sharing his copy of Paul Merrill's monograph,
``Lines of the Chemical Elements In Astronomical Spectra.''
This research has made use of NASA's
Astrophysics Data System Bibliographic Services,
the arXiv pre-print server operated by Cornell University,
the SIMBAD and VizieR databases hosted by the
Strasbourg Astronomical Data Center,
the Atomic Spectra Database hosted by
the National Institute of Standards and Technology, 
the Multimission Archive at the Space Telescope Science Institute,
the ESO Science Archive Facility,
and the Keck Observatory Archive.
IRAF is distributed by the National Optical Astronomy Observatories,
which are operated by the Association of Universities for Research
in Astronomy, Inc., under cooperative agreement with the National
Science Foundation.
I.U.R.\ is supported by the Carnegie Institution of Washington 
through the Carnegie Observatories Fellowship.
J.E.L.\ acknowledges support from NASA Grant NNX10AN93G.

{\it Facilities:} 
\facility{HST (STIS)},
\facility{ESO:3.6m (HARPS)},
\facility{VLT:Kueyen (UVES)},
\facility{Keck:I (HIRES)}

\end{document}

%% file: tab1.tex
\begin{deluxetable*}{ccccccc}
\tablecaption{Transition probabilities for Cu~\textsc{ii} lines of interest
\label{cutab}}
\tablewidth{0pt}
\tabletypesize{\scriptsize}
\tablehead{
\colhead{Wavelength\tablenotemark{a}} &
\colhead{E$_{\rm upper}$} &
\colhead{J$_{\rm upper}$} &
\colhead{E$_{\rm lower}$} &
\colhead{J$_{\rm lower}$} &
\colhead{A} &
\colhead{$\log(gf)$} \\
\colhead{(\AA)} &
\colhead{(cm$^{-1}$)} &
\colhead{} &
\colhead{(cm$^{-1}$)} &
\colhead{} &
\colhead{(10$^{6}$ s$^{-1}$)} &
\colhead{} }
\startdata
2135.9808 & 68730.893 & 4 & 21928.754 & 3 & 459$\pm$8  & $+$0.45 \\
2126.0443 & 69867.983 & 2 & 22847.131 & 2 & 172        & $-$0.23 \\
2112.1003 & 73595.813 & 1 & 26264.568 & 2 & 384        & $-$0.11 \\
2104.7963 & 71493.853 & 2 & 23998.381 & 1 &  92$\pm$7  & $-$0.51 \\
2054.9790 & 71493.853 & 2 & 22847.131 & 2 & 162$\pm$12 & $-$0.29 \\
2037.1270 & 71920.102 & 3 & 22847.131 & 2 & 134$\pm$10 & $-$0.23 \\
1979.9565 & 73353.292 & 2 & 22847.131 & 2 &  88$\pm$5  & $-$0.59 \\
\enddata
\tablenotetext{a}{
Air wavelengths are given for $\lambda >$~2000\,\AA\ 
and vacuum values below.
}
\end{deluxetable*}

%% file: tab2.tex
\begin{deluxetable*}{cccccccc}
\tablecaption{Transition probabilities for Zn~\textsc{i} and Zn~\textsc{ii} lines of interest
\label{zntab}}
\tablewidth{0pt}
\tabletypesize{\scriptsize}
\tablehead{
\colhead{Air wavelength} &
\colhead{E$_{\rm upper}$} &
\colhead{J$_{\rm upper}$} &
\colhead{E$_{\rm lower}$} &
\colhead{J$_{\rm lower}$} &
\colhead{A} &
\colhead{$\log(gf)$} &
\colhead{Spectrum} \\
\colhead{(\AA)} &
\colhead{(cm$^{-1}$)} &
\colhead{} &
\colhead{(cm$^{-1}$)} &
\colhead{} &
\colhead{(10$^{6}$ s$^{-1}$)} &
\colhead{} &
\colhead{} }
\startdata
4810.5321 & 53672.240 & 1 & 32890.327 & 2 &  67$\pm$3        & $-$0.15 & Zn~\textsc{i} \\
4722.1569 & 53672.240 & 1 & 32501.399 & 1 &  43$\pm$2        & $-$0.37 & Zn~\textsc{i} \\
4680.1362 & 53672.240 & 1 & 32311.319 & 0 & 15.0$\pm$0.8     & $-$0.85 & Zn~\textsc{i} \\
3345.9353 & 62768.747 & 1 & 32890.327 & 2 &   4$\pm$0.2      & $-$1.66\tablenotemark{a} & Zn~\textsc{i} \\
3345.5695 & 62772.014 & 2 & 32890.327 & 2 &  38$\pm$2        & $-$0.51 & Zn~\textsc{i} \\
3345.0134 & 62776.981 & 3 & 32890.327 & 2 & 156$\pm$8        & $+$0.26 & Zn~\textsc{i} \\
3302.9394 & 62768.747 & 1 & 32501.399 & 1 &  66$\pm$3        & $-$0.48 & Zn~\textsc{i} \\
3302.5829 & 62772.014 & 2 & 32501.399 & 1 & 118$\pm$6        & $-$0.02 & Zn~\textsc{i} \\
3282.3256 & 62768.747 & 1 & 32311.319 & 0 &  90$\pm$5        & $-$0.36 & Zn~\textsc{i} \\
3075.8970 & 32501.399 & 1 &     0.000 & 0 & 0.0329           & $-$3.85 & Zn~\textsc{i} \\
2062.0012 & 48481.077 & 0.5 & 0.000 & 0.5 & 400$\pm$32       & $-$0.29 & Zn~\textsc{ii}\\
\enddata
\tablenotetext{a}{
The $\log(gf)$ value has been adjusted by $+$0.04~dex to correct 
a truncation in the Einstein A value from \citet{kerkhoff80}.
}
\end{deluxetable*}

%% file: tab3.tex
\begin{deluxetable*}{ccccccc}
\tablecaption{Transition probabilities for As~\textsc{i} lines of interest
\label{astab}}
\tablewidth{0pt}
\tabletypesize{\scriptsize}
\tablehead{
\colhead{Vacuum wavelength} &
\colhead{E$_{\rm upper}$} &
\colhead{J$_{\rm upper}$} &
\colhead{E$_{\rm lower}$} &
\colhead{J$_{\rm lower}$} &
\colhead{A} &
\colhead{$\log(gf)$} \\
\colhead{(\AA)} &
\colhead{(cm$^{-1}$)} &
\colhead{} &
\colhead{(cm$^{-1}$)} &
\colhead{} &
\colhead{(10$^{6}$ s$^{-1}$)} &
\colhead{} }
\startdata
1990.3552 & 60834.954 & 1.5 & 10592.666 & 1.5 & 220        & $-$0.28 \\
1972.6240 & 50693.897 & 0.5 &     0.000 & 1.5 & 202$\pm$22 & $-$0.63 \\
1937.5942 & 51610.393 & 1.5 &     0.000 & 1.5 & 219$\pm$25 & $-$0.31 \\
1890.4286 & 52898.056 & 2.5 &     0.000 & 1.5 & 200        & $-$0.19 \\
\enddata
\end{deluxetable*}

%% file: tab4-stub.tex
\begin{deluxetable*}{ccccccc}
\tablecaption{Hyperfine structure line component patterns for $^{75}$As~\textsc{i}
\label{ashfstab}}
\tablewidth{0pt}
\tabletypesize{\scriptsize}
\tablehead{
\colhead{Wavenumber} &
\colhead{$\lambda_{\rm vac}$} &
\colhead{F$_{\rm upper}$} &
\colhead{F$_{\rm lower}$} &
\colhead{Component Position} &
\colhead{Component Position} &
\colhead{Strength} \\
\colhead{(cm$^{-1}$)} &
\colhead{(\AA)} &
\colhead{} &
\colhead{} &
\colhead{(cm$^{-1}$)} &
\colhead{(\AA)} &
\colhead{} }
\startdata
50242.288 & 1990.3552 & 3 & 3 &    0.01800 & $-$0.000713 & 0.35000 \\
50242.288 & 1990.3552 & 3 & 2 &    0.06600 & $-$0.002615 & 0.08750 \\
50242.288 & 1990.3552 & 2 & 3 & $-$0.05400 &    0.002139 & 0.08750 \\
\enddata
\tablecomments{
The line component patterns are computed from HFS constants of
\citet{pendlebury64} and \citet{bouazza87}.
Center-of-gravity wavenumbers and vacuum wavelengths, 
$\lambda_{\rm vac}$, from \citet{howard85}, are given
with component positions relative to those values.
Strengths are normalized to sum to 1.
Table~\ref{ashfstab} is available in its entirety via the
link to the machine-readable table above.}
\end{deluxetable*}

%% file: tab5.tex
\begin{deluxetable*}{ccccccc}
\tablecaption{Transition probabilities for Se~\textsc{i} lines of interest
\label{setab}}
\tablewidth{0pt}
\tabletypesize{\scriptsize}
\tablehead{
\colhead{Wavelength\tablenotemark{a}} &
\colhead{E$_{\rm upper}$} &
\colhead{J$_{\rm upper}$} &
\colhead{E$_{\rm lower}$} &
\colhead{J$_{\rm lower}$} &
\colhead{A} &
\colhead{$\log(gf)$} \\
\colhead{(\AA)} &
\colhead{(cm$^{-1}$)} &
\colhead{} &
\colhead{(cm$^{-1}$)} &
\colhead{} &
\colhead{(10$^{6}$ s$^{-1}$)} &
\colhead{} }
\startdata
2074.7841 & 48182.420 & 2 &    0.00  & 2 & 1.70$\pm$0.10 & $-$2.26 \\
2062.7789 & 50997.161 & 1 & 2534.36  & 0 & 33$\pm$6      & $-$1.20 \\
2039.8420 & 50997.161 & 1 & 1989.497 & 1 & 98$\pm$17     & $-$0.74 \\
1960.8935 & 50997.161 & 1 &    0.00  & 2 & 213$\pm$37    & $-$0.43 \\
\enddata
\tablenotetext{a}{
Air wavelengths are given for $\lambda >$~2000\,\AA\ and vacuum values below.
}
\end{deluxetable*}

%% file: tab6.tex
\begin{deluxetable*}{ccccccc}
\tablecaption{Transition probabilities for Mo~\textsc{ii} lines of interest
\label{motab}}
\tablewidth{0pt}
\tabletypesize{\scriptsize}
\tablehead{
\colhead{Air wavelength} &
\colhead{E$_{\rm upper}$} &
\colhead{J$_{\rm upper}$} &
\colhead{E$_{\rm lower}$} &
\colhead{J$_{\rm lower}$} &
\colhead{A} &
\colhead{$\log(gf)$} \\
\colhead{(\AA)} &
\colhead{(cm$^{-1}$)} &
\colhead{} &
\colhead{(cm$^{-1}$)} &
\colhead{} &
\colhead{(10$^{6}$ s$^{-1}$)} &
\colhead{} }
\startdata
2045.9729 & 48860.829 & 2.5 & 0.000 & 2.5 & 118$\pm$8  & $-$0.35 \\
2038.4522 & 49041.073 & 1.5 & 0.000 & 2.5 & 184$\pm$18 & $-$0.34 \\
2020.3139 & 49481.300 & 3.5 & 0.000 & 2.5 & 215$\pm$20 & $+$0.02 \\
2015.1091 & 49609.086 & 2.5 & 0.000 & 2.5 &  88$\pm$6  & $-$0.49 \\
\enddata
\end{deluxetable*}

%% file: tab7.tex
\begin{deluxetable*}{ccccccc}
\tablecaption{Transition probabilities for Cd~\textsc{ii} lines of interest
\label{cdtab}}
\tablewidth{0pt}
\tabletypesize{\scriptsize}
\tablehead{
\colhead{Air wavelength} &
\colhead{E$_{\rm upper}$} &
\colhead{J$_{\rm upper}$} &
\colhead{E$_{\rm lower}$} &
\colhead{J$_{\rm lower}$} &
\colhead{A} &
\colhead{$\log(gf)$} \\
\colhead{(\AA)} &
\colhead{(cm$^{-1}$)} &
\colhead{} &
\colhead{(cm$^{-1}$)} &
\colhead{} &
\colhead{(10$^{6}$ s$^{-1}$)} &
\colhead{} }
\startdata
2265.0145 & 44136.173 & 0.5 & 0.00 & 0.5 & 317.7$\pm$1.1 & $-$0.311 \\
2144.3943 & 46618.532 & 1.5 & 0.00 & 0.5 & 377.8$\pm$1.4 & $+$0.018 \\
\enddata
\end{deluxetable*}

%% file: tab8-stub.tex
\begin{deluxetable*}{cccccccc}
\tablecaption{Hyperfine structure and isotopic line component patterns for 
Cd~\textsc{ii} lines
\label{cdhfstab}}
\tablewidth{0pt}
\tabletypesize{\scriptsize}
\tablehead{
\colhead{Wavenumber} &
\colhead{$\lambda_{\rm air}$} &
\colhead{F$_{\rm upper}$} &
\colhead{F$_{\rm lower}$} &
\colhead{Component Position} &
\colhead{Component Position} &
\colhead{Strength} &
\colhead{Isotope} \\
\colhead{(cm$^{-1}$)} &
\colhead{(\AA)} &
\colhead{} &
\colhead{} &
\colhead{(cm$^{-1}$)} &
\colhead{(\AA)} &
\colhead{} &
\colhead{} }
\startdata
44136.173 & 2265.0145 & 0.5 & 0.5 &    0.07759 & $-$0.003982 & 0.01250 & 106 \\
44136.173 & 2265.0145 & 0.5 & 0.5 &    0.05164 & $-$0.002650 & 0.00890 & 108 \\
44136.173 & 2265.0145 & 0.5 & 0.5 &    0.02684 & $-$0.001378 & 0.12490 & 110 \\
\enddata
\tablecomments{
Center-of-gravity wavenumbers and air wavelengths, 
$\lambda_{\rm air}$, from \citet{burns56}, are given
with component positions relative to those values.
Strengths are normalized to sum to 1.
Table~\ref{cdhfstab} is available in its entirety via the
link to the machine-readable table above.}
\end{deluxetable*}

%% file: tab9.tex
\begin{deluxetable*}{ccccccc}
\tablecaption{Transition probabilities for Yb~\textsc{ii} lines of interest
\label{ybtab}}
\tablewidth{0pt}
\tabletypesize{\scriptsize}
\tablehead{
\colhead{Air wavelength} &
\colhead{E$_{\rm upper}$} &
\colhead{J$_{\rm upper}$} &
\colhead{E$_{\rm lower}$} &
\colhead{J$_{\rm lower}$} &
\colhead{A} &
\colhead{$\log(gf)$} \\
\colhead{(\AA)} &
\colhead{(cm$^{-1}$)} &
\colhead{} &
\colhead{(cm$^{-1}$)} &
\colhead{} &
\colhead{(10$^{6}$ s$^{-1}$)} &
\colhead{} }
\startdata
2126.741 & 47005.46 & 1.5 & 0.00 & 0.5 & 50.2$\pm$1.3 & $-$0.87 \\
2116.675 & 47228.96 & 0.5 & 0.00 & 0.5 & 34.1         & $-$1.34 \\
\enddata
\end{deluxetable*}

%% file: tab10-stub.tex
\begin{deluxetable*}{cccccccc}
\tablecaption{Hyperfine structure and isotopic line component patterns for 
Yb~\textsc{ii} lines
\label{ybhfstab}}
\tablewidth{0pt}
\tabletypesize{\scriptsize}
\tablehead{
\colhead{Wavenumber} &
\colhead{$\lambda_{\rm air}$} &
\colhead{F$_{\rm upper}$} &
\colhead{F$_{\rm lower}$} &
\colhead{Component Position} &
\colhead{Component Position} &
\colhead{Strength} &
\colhead{Isotope} \\
\colhead{(cm$^{-1}$)} &
\colhead{(\AA)} &
\colhead{} &
\colhead{} &
\colhead{(cm$^{-1}$)} &
\colhead{(\AA)} &
\colhead{} &
\colhead{} }
\startdata
47005.46 & 2126.741 & 1.5 & 0.5 & $-$0.07332 &    0.003318 & 0.00130 & 168 \\
47005.46 & 2126.741 & 1.5 & 0.5 & $-$0.04850 &    0.002195 & 0.03040 & 170 \\
47005.46 & 2126.741 & 1.5 & 0.5 & $-$0.01332 &    0.000603 & 0.21830 & 172 \\
\enddata
\tablecomments{
Center-of-gravity wavenumbers and air wavelengths, 
$\lambda_{\rm air}$, are from \citet{martin78},
and the index of air \citep{peck72} 
with component positions relative to those values.
Strengths are normalized to sum to 1.
Table~\ref{ybhfstab} is available in its entirety via the
link to the machine-readable table above.
The hfs A values of the upper levels are neglected for the odd isotopes.
}
\end{deluxetable*}

%% file: tab11.tex
\begin{deluxetable*}{ccccccc}
\tablecaption{Transition probabilities for Pt~\textsc{i} lines of interest
\label{pttab}}
\tablewidth{0pt}
\tabletypesize{\scriptsize}
\tablehead{
\colhead{Air wavelength} &
\colhead{E$_{\rm upper}$} &
\colhead{J$_{\rm upper}$} &
\colhead{E$_{\rm lower}$} &
\colhead{J$_{\rm lower}$} &
\colhead{A} &
\colhead{$\log(gf)$} \\
\colhead{(\AA)} &
\colhead{(cm$^{-1}$)} &
\colhead{} &
\colhead{(cm$^{-1}$)} &
\colhead{} &
\colhead{(10$^{6}$ s$^{-1}$)} &
\colhead{} }
\startdata
2103.3441 & 48351.940 & 4 & 823.678 & 4 & 119$\pm$6    & $-$0.15 \\
2067.5090 & 48351.940 & 4 &   0.000 & 3 & 42.0$\pm$2.7 & $-$0.62 \\
2049.3914 & 48779.337 & 3 &   0.000 & 3 & 235$\pm$29   & $+$0.02 \\
\enddata
\end{deluxetable*}

%% file: tab12-stub.tex
\begin{deluxetable*}{cccccccc}
\tablecaption{Hyperfine structure and S.S.\ isotopic line component pattern 
of the Hg~\textsc{ii} 1942\,\AA\ line
\label{hghfstab}}
\tablewidth{0pt}
\tabletypesize{\scriptsize}
\tablehead{
\colhead{Wavenumber} &
\colhead{$\lambda_{\rm vacuum}$} &
\colhead{F$_{\rm upper}$} &
\colhead{F$_{\rm lower}$} &
\colhead{Component Position} &
\colhead{Component Position} &
\colhead{Strength} &
\colhead{Isotope} \\
\colhead{(cm$^{-1}$)} &
\colhead{(\AA)} &
\colhead{} &
\colhead{} &
\colhead{(cm$^{-1}$)} &
\colhead{(\AA)} &
\colhead{} &
\colhead{} }
\startdata
51486.070 & 1942.2729 & 0.5 & 0.5 &  0.44405 & -0.016751 & 0.00150 & 196 \\
51486.070 & 1942.2729 & 0.5 & 0.5 &  0.24705 & -0.009320 & 0.09970 & 198 \\
51486.070 & 1942.2729 & 0.5 & 0.5 &  0.05005 & -0.001888 & 0.23100 & 200 \\
\enddata
\tablecomments{
Center-of-gravity wavenumbers and vacuum wavelengths, 
$\lambda_{\rm vacuum}$, are for a S.S.\ isotopic mix
from \citet{sansonetti01}
with component positions relative to those values.
Strengths are normalized to sum to 1.
Table~\ref{hghfstab} is available in its entirety via the
link to the machine-readable table above.
}
\end{deluxetable*}

%% file: tab13-stub.tex
\begin{deluxetable*}{cccccc}
\tablecaption{Equivalent Width Measurements
\label{ewtab}}
\tablewidth{0pt}
\tabletypesize{\scriptsize}
\tablehead{
\colhead{Wavelength (\AA)} &
\colhead{Species} &
\colhead{E.P. (eV)} &
\colhead{log($gf$)} &
\colhead{EW (m\AA)} &
\colhead{Instrument} }
\startdata
   5682.63                  & Na~\textsc{i} &   2.100  &  -0.706   &    5.3 & HARPS \\
   5688.20                  & Na~\textsc{i} &   2.100  &  -0.452   &   14.3 & HARPS \\
   4167.27                  & Mg~\textsc{i} &   4.346  &  -0.710   &   58.1 & HARPS \\
\enddata
\tablecomments{
Table~\ref{ewtab} is available in its entirety in the online version
of the journal.
A portion is shown here to illustrate its form and content.}              
\tablenotetext{a}{EW measured but line rejected from abundance 
computation; see Section~\ref{analysis}.}
\end{deluxetable*}

%% file: tab14-mystub.tex
\begin{deluxetable*}{ccccccc}
\tablecaption{Wavelengths, excitation potentials, $\log(gf)$ values, 
and derived abundances
\label{linetab}}
\tablewidth{0pt}
\tabletypesize{\scriptsize}
\tablehead{
\colhead{Species} &
\colhead{Wavelength (\AA)\tablenotemark{a}} &
\colhead{E.P. (eV)} &
\colhead{$\log(gf)$} &
\colhead{$\log \epsilon$} &
\colhead{Instrument} &
\colhead{Ref.} }
\startdata
Be~\textsc{ii} &    3130.42  & 0.00     & -0.18          & -0.35 &  UVES   &  1  \\ 
Be~\textsc{ii} &    3131.07  & 0.00     & -0.48          & -0.38 &  UVES   &  1  \\ 
C~(CH)         & 4290--4325  & \nodata  & \nodata        & 6.51  &  HARPS  &  2  \\ 
\enddata                                       
\tablecomments{Table~\ref{linetab} will be available in its entirety
in the final edition of the journal. 
Please contact the first author for an advance copy.
}
\tablenotetext{a}{Air wavelengths are given for $\lambda >$~2000\,\AA\
and vacuum values below}                       
\tablerefs{
(1) \citealt{fuhr09};
(2) B.\ Plez, 2007, private communication;
(3) \citealt{kurucz95};
(4) \citealt{chang90};
(5) \citealt{aldenius09};
(6) \citealt{lawler89}, using hfs from \citealt{kurucz95};
(7) \citealt{blackwell82a,blackwell82b}, increased by 0.056~dex according to \citealt{grevesse89};
(8) \citealt{pickering01}, with corrections given in \citealt{pickering02};
(9) \citealt{doerr85}, using hfs from \citealt{kurucz95};
(10) \citealt{biemont89};
(11) \citealt{sobeck07};
(12) \citealt{nilsson06};
(13) \citealt{denhartog11} for both log($gf$) value and hfs;
(14) \citealt{obrian91};
(15) \citealt{melendez09};
(16) \citealt{cardon82}, using hfs from \citealt{kurucz95};
(17) \citealt{nitz99}, using hfs from \citealt{kurucz95};
(18) \citealt{blackwell89};
(19) \citealt{bielski75}, using hfs/IS from J.S.\ Sobeck et al.\ in prep.;
(20) this study;
(21) \citealt{kerkhoff80};
(22) \citealt{bergeson93};
(23) \citealt{morton00};
(24) \citealt{holmgren75}, using hfs from this study;
(25) \citealt{parkinson76};
(26) \citealt{biemont11};
(27) \citealt{malcheva06};
(28) \citealt{ljung06};
(29) \citealt{whaling88};
(30) \citealt{sikstrom01};
(31) \citealt{wickliffe94};
(32) \citealt{xu06};
(33) \citealt{fuhr09}, using hfs/IS from \citealt{hansen12};
(34) \citealt{roederer12a};
(35) \citealt{fuhr09}, using hfs/IS from \citealt{mcwilliam98} when available;
(36) \citealt{lawler01a}, using hfs from \citealt{ivans06} when available;
(37) \citealt{lawler09};
(38) \citealt{denhartog03}, using hfs/IS from \citealt{roederer08} when available;
(39) \citealt{lawler06}, using hfs/IS from \citealt{roederer08} when available;
(40) \citealt{lawler01b}, using hfs/IS from \citealt{ivans06};
(41) \citealt{denhartog06};
(42) \citealt{wickliffe00};
(43) \citealt{lawler08};
(44) $\log(gf)$ value from the Database on Rare Earths At Mons University (DREAM), using hfs/IS from this study;
(45) \citealt{kedzierski10}, using hfs/IS from this study;
(46) \citealt{sneden09} for both $\log(gf)$ and hfs/IS;
(47) \citealt{ivarsson04};
(48) \citealt{denhartog05};
(49) \citealt{biemont00}, using hfs/IS from \citealt{roederer12b}
}
\end{deluxetable*}

%% file: tab15.tex
\begin{deluxetable*}{cccccccc}
\tablecaption{Mean abundances in HD 160617
\label{abundtab}}
\tablewidth{0pt}
\tabletypesize{\scriptsize}
\tablehead{
\colhead{$Z$} &
\colhead{Species} &
\colhead{$\log \epsilon$ S.S.\tablenotemark{a}} &
\colhead{$\log \epsilon$} &
\colhead{[X/Fe]\tablenotemark{b}} &
\colhead{N$_{\rm lines}$} &
\colhead{$\sigma_{\rm internal}$} &
\colhead{$\sigma_{\rm total}$} }
\startdata
4      & Be~\textsc{ii}   & 1.38    & $-$0.36    & $+$0.03    & 2       & 0.07    & 0.30    \\ 
6      & C~(CH)           & 8.43    & $+$6.61    & $+$0.10    & \nodata & 0.15    & 0.25    \\
7      & N~(NH)           & 7.83    & $+$6.51    & $+$0.60    & \nodata & 0.15    & 0.25    \\
8      & O~(OH)           & 8.69    & $+$7.22    & $+$0.45    & \nodata & 0.15    & 0.25    \\
11     & Na~\textsc{i}    & 6.24    & $+$4.58    & $+$0.26    & 2       & 0.07    & 0.20    \\
12     & Mg~\textsc{i}    & 7.60    & $+$5.88    & $+$0.20    & 4       & 0.23    & 0.30    \\
14     & Si~\textsc{i}    & 7.51    & $+$5.91    & $+$0.32    & 3       & 0.07    & 0.20    \\
20     & Ca~\textsc{i}    & 6.34    & $+$4.78    & $+$0.36    & 13      & 0.09    & 0.21    \\
21     & Sc~\textsc{ii}   & 3.15    & $+$1.45    & $+$0.07    & 8       & 0.15    & 0.33    \\
22     & Ti~\textsc{i}    & 4.95    & $+$3.31    & $+$0.28    & 16      & 0.04    & 0.19    \\
22     & Ti~\textsc{ii}   & 4.95    & $+$3.47    & $+$0.29    & 28      & 0.08    & 0.30    \\
23     & V~\textsc{i}     & 3.93    & $+$2.08    & $+$0.07    & 1       & 0.10    & 0.33    \\
23     & V~\textsc{ii}    & 3.93    & $+$2.34    & $+$0.18    & 2       & 0.07    & 0.30    \\
24     & Cr~\textsc{i}    & 5.64    & $+$3.62    & $-$0.10    & 13      & 0.05    & 0.20    \\
24     & Cr~\textsc{ii}   & 5.64    & $+$4.08    & $+$0.21    & 3       & 0.06    & 0.30    \\
25     & Mn~\textsc{i}    & 5.43    & $+$3.19    & $-$0.32    & 5       & 0.04    & 0.19    \\
25     & Mn~\textsc{ii}   & 5.43    & $+$3.37    & $-$0.29    & 6       & 0.04    & 0.29    \\
26     & Fe~\textsc{i}    & 7.50    & $+$5.58    & $-$1.92\tablenotemark{c} & 119  & 0.06 & 0.20    \\
26     & Fe~\textsc{ii}   & 7.50    & $+$5.73    & $-$1.77\tablenotemark{d} & 11   & 0.05 & 0.29    \\
27     & Co~\textsc{i}    & 4.99    & $+$3.07    & $+$0.00    & 7       & 0.13    & 0.23    \\
28     & Ni~\textsc{i}    & 6.22    & $+$4.30    & $+$0.00    & 14      & 0.05    & 0.20    \\
29     & Cu~\textsc{i}    & 4.19    & $+$1.40    & $-$0.87    & 2       & 0.07    & 0.20    \\
29     & Cu~\textsc{ii}   & 4.19    & $+$1.84    & $-$0.58    & 6       & 0.05    & 0.29    \\
30     & Zn~\textsc{i}    & 4.56    & $+$2.76    & $+$0.12    & 6       & 0.05    & 0.20    \\
30     & Zn~\textsc{ii}   & 4.56    & $+$2.94    & $+$0.15    & 1       & 0.10    & 0.31    \\
32     & Ge~\textsc{i}    & 3.65    & $<+$1.20   & $<-$0.53   & \nodata & \nodata & \nodata \\
33     & As~\textsc{i}    & 2.30    & $+$0.60    & $+$0.22    & 3       & 0.37    & 0.42    \\
34     & Se~\textsc{i}    & 3.34    & $+$1.56    & $+$0.14    & 3       & 0.08    & 0.21    \\
38     & Sr~\textsc{i}    & 2.87    & $+$0.85    & $-$0.10    & 1       & 0.10    & 0.21    \\
38     & Sr~\textsc{ii}   & 2.87    & $+$1.22    & $+$0.12    & 3       & 0.09    & 0.30    \\
39     & Y~\textsc{ii}    & 2.21    & $+$0.25    & $-$0.19    & 11      & 0.08    & 0.30    \\
40     & Zr~\textsc{ii}   & 2.58    & $+$1.05    & $+$0.24    & 20      & 0.07    & 0.30    \\
42     & Mo~\textsc{i}    & 1.88    & $+$0.58    & $+$0.62    & 1       & 0.10    & 0.22    \\
42     & Mo~\textsc{ii}   & 1.88    & $+$0.61    & $+$0.50    & 4       & 0.09    & 0.30    \\
44     & Ru~\textsc{i}    & 1.75    & $+$0.59    & $+$0.76    & 1       & 0.10    & 0.22    \\
46     & Pd~\textsc{i}    & 1.65    & $-$0.07    & $+$0.20    & 1       & 0.10    & 0.22    \\
47     & Ag~\textsc{i}    & 1.20    & $-$0.89    & $-$0.17    & 1       & 0.20    & 0.28    \\
48     & Cd~\textsc{ii}   & 1.71    & $-$0.03    & $+$0.03    & 1       & 0.10    & 0.31    \\
52     & Te~\textsc{i}    & 2.18    & $+$0.67    & $+$0.41    & 1       & 0.20    & 0.32    \\
56     & Ba~\textsc{ii}   & 2.18    & $+$0.44    & $+$0.03    & 2       & 0.07    & 0.30    \\
57     & La~\textsc{ii}   & 1.10    & $-$0.36    & $+$0.31    & 4       & 0.07    & 0.30    \\
58     & Ce~\textsc{ii}   & 1.58    & $+$0.06    & $+$0.25    & 9       & 0.09    & 0.30    \\
60     & Nd~\textsc{ii}   & 1.42    & $-$0.03    & $+$0.32    & 4       & 0.09    & 0.30    \\
62     & Sm~\textsc{ii}   & 0.96    & $-$0.45    & $+$0.36    & 3       & 0.15    & 0.33    \\
63     & Eu~\textsc{ii}   & 0.52    & $-$0.81    & $+$0.44    & 4       & 0.05    & 0.29    \\
64     & Gd~\textsc{ii}   & 1.07    & $-$0.30    & $+$0.40    & 3       & 0.06    & 0.30    \\
66     & Dy~\textsc{ii}   & 1.10    & $-$0.20    & $+$0.47    & 6       & 0.04    & 0.29    \\
68     & Er~\textsc{ii}   & 0.92    & $-$0.35    & $+$0.50    & 3       & 0.06    & 0.30    \\
70     & Yb~\textsc{ii}   & 0.92    & $-$0.42    & $+$0.43    & 3       & 0.15    & 0.33    \\
76     & Os~\textsc{ii}   & 1.40    & $-$0.12    & $+$0.25    & 1       & 0.20    & 0.36    \\
77     & Ir~\textsc{ii}   & 1.38    & $+$0.35    & $+$0.74    & 1       & 0.10    & 0.31    \\
78     & Pt~\textsc{i}    & 1.62    & $+$0.44    & $+$0.74    & 3       & 0.10    & 0.21    \\
80     & Hg~\textsc{ii}   & 1.17    & $<+$0.40   & $<+$1.00   & \nodata & \nodata & \nodata \\
82     & Pb~\textsc{i}    & 2.04    & $<+$0.90   & $<+$0.78   & \nodata & \nodata & \nodata \\
\enddata
\tablenotetext{a}{Solar abundances from \citet{asplund09}}              
\tablenotetext{b}{Abundance ratios for 
neutrals are computed relative to Fe~\textsc{i} and
abundance ratios for ions are computed
relative to Fe~\textsc{ii}}
\tablenotetext{c}{[Fe~\textsc{i}/H]}
\tablenotetext{d}{[Fe~\textsc{ii}/H]}
\end{deluxetable*}